\documentclass[preprint,superscriptaddress,nofootinbib,12pt]{revtex4-1}
\usepackage{array,multirow} 
\usepackage{amssymb, amsmath, graphicx,float}
\usepackage{color}

\pdfoutput=1

\date{\today}

\def\lag{\mathcal{L}}
\def\lagint{\lag_\text{int}}
\def\op{\mathcal{O}}
\def\oprel{\mathcal{O}^\text{rel}}
\def\chibar{\bar{\chi}}
\def\Nbar{\bar{N}}

\def\W{\tilde{W}}
\def\sp{\langle {S}_p \rangle }
\def\sn{\langle {S}_n \rangle }

\def\Lp{\langle {L}_p \rangle }
\def\Ln{\langle {L}_n \rangle }
\def\LN{\langle {L}_N \rangle }

\def\mag{\tilde{\mu}}

\def\beq{\begin{equation}}
\def\eeq{\end{equation}}

\def\bit{\begin{itemize}}
\def\eit{\end{itemize}}
\def\ben{\begin{enumerate}}
\def\een{\end{enumerate}}

\begin{document}

\title{On the Effect of Nuclear Response Functions in Dark Matter Direct Detection}

\author{Moira I. Gresham}
\affiliation{Whitman College, Walla Walla, WA 99362, USA}
\author{Kathryn M. Zurek}
\affiliation{Michigan Center for Theoretical Physics, University of Michigan, Ann Arbor, MI 48109, USA}

\begin{abstract}

We examine the effect of nuclear response functions, as laid out in \cite{Fitzpatrick:2012ix}, on dark matter (DM) direct detection in the context of well-motivated UV completions, including electric and magnetic dipoles, anapole, spin-orbit, and pseudoscalar-mediated DM.  Together, these encompass five of the six nuclear responses extracted from the non-relativistic effective theory of \cite{Fitzpatrick:2012ix} (with the sixth difficult to UV complete), with two of the six combinations  corresponding to standard spin-independent and -dependent responses. For constraints from existing direct detection experiments, we find that only the COUPP constraint, due to its heavy iodine target with large angular momentum and an unpaired spin, and its large energy range sensitivity, is substantially modified by the new responses compared to what would be inferred using the standard form factors to model the energy dependence of the response.
 For heavy targets such as xenon and germanium, the behavior of the new nuclear responses as recoil energy increases can be substantially different than that of the standard responses, but this has almost no impact on the constraints derived from experiments such as LUX, XENON100 and CDMS since the maximum nuclear recoil energy detected in these experiments is relatively low.  We simulate mock data for 80 and 250 GeV DM candidates utilizing the new nuclear responses to highlight how they might affect a putative signal, and find the new responses are most important for momentum-suppressed interactions such as the magnetic dipole or pseudoscalar-mediated interaction when the target is relatively heavy (such as xenon and iodine). 

\end{abstract}
\preprint{MCTP-14-01}

\maketitle
\tableofcontents

\section{Introduction}\label{sec: intro}

Detection of Dark Matter (DM) directly via its scattering off of nuclei in a radiopure underground detector is  currently one of the best probes of the DM sector. 
A host of experiments, such as XENON100, LUX, CDMS, and COUPP \cite{Aprile:2012nq,Akerib:2013tjd,Ahmed:2009zw,Agnese:2013lua,Behnke:2012ys}, are cutting into the parameter space for weakly interacting massive particles (WIMPs), while possible signals (now highly constrained) \cite{Bernabei:2010mq,Angloher:2011uu,Aalseth:2012if,Agnese:2013rvf} have raised interest.   Extracting meaningful bounds on the nature of the DM from these experiments requires theoretical inputs. The inputs include the type of  DM-nucleus coupling as well as the nuclear response to the DM interaction, which depend on the new physics mediating the scattering and known nuclear physics, respectively.

For interactions through ordinary spin-independent and -dependent operators, the nature of the DM interactions, the effect of the astrophysics on the scattering rates, and the nuclear form factors have been well studied and the associated uncertainties understood. For classic and recent reviews see \cite{Engel:1992bf,Jungman:1995df,Lewin:1995rx,Bertone:2004pz,Bednyakov:2006ux,Ellis:2008hf,Feng:2010gw,Green:2011bv,Peter:2013aha,Zurek:2013wia,Cushman:2013zza}. On the other hand, while most of the literature assumes that DM couples to nuclei primarily through ordinary spin-dependent or -independent interactions, a much broader array of possibilities is well motivated.    These occur, {\em e.g.}, when the DM couples through a moment  ({\em esp.}~magnetic dipole or anapole), is subject to dark forces, is composite, and/or the mediating particle is a pseudoscalar \cite{Pospelov:2000bq,Bagnasco:1993st,Sigurdson:2004zp,Chang:2009yt,Feldstein:2009tr,Fitzpatrick:2010br,Banks:2010eh,An:2010kc,Barger:2010gv,Dienes:2013xya}.   An effective operator description in the context of non-relativistic effective theory encompassing scattering models with nonstandard interactions was explored in \cite{Fan:2010gt,Fitzpatrick:2012ix,Fitzpatrick:2012ib,DelNobile:2013sia,Anand:2013yka}.  In particular, \cite{Fitzpatrick:2012ix} pointed out that a number of the operators in this non-relativistic description involve novel nuclear responses, which had not been included in previous analyses.

In the standard case (spin-independent or -dependent interactions), the nuclear response is encoded at zero momentum transfer in the mass number, $A$, and total charge, $Z$, for spin-independent interactions, or in total angular momentum, $J$, and average expected nucleon spin, $\langle S_p \rangle$, $\langle S_n \rangle$, for spin-dependent interactions.  Nuclear form factors describe the change of nuclear response with increasing momentum transfer because of the composite nature of nuclei. The standard spin-independent and -dependent form factors can be looked up in tables in the literature for various elements---for example, in the references listed in Table~\ref{tab: spins etc}.  (In the case of spin-independent interactions, the Helm form factor is usually used.) When new types of nuclear responses are excited due to nonstandard interactions between the DM and nuclei, however, different form factors than the standard spin-independent and -dependent ones should be employed. By selecting the relevant non-relativistic building blocks for DM scattering, \cite{Fitzpatrick:2012ix} was able to elucidate the relevant nuclear responses for nonstandard DM interactions;  they also showed how to map their non-relativistic results onto relativistic operators.

In particular, Ref.~\cite{Fitzpatrick:2012ix} showed that there are six independent types of nuclear responses that can be relevant for DM scattering---rather than just the two (spin-independent and -dependent) standardly considered.  These arise when the relative DM or nucleon velocities or momentum transfer is intertwined with the DM or nucleon spin in the underlying DM-nucleon interaction.   These responses, along with their zero momentum limit, are shown in Table~\ref{tab: responses}.  To make contact with more familiar language, the standard spin-independent nuclear response is $M$ (which closely mimics the Helm form factor), 
while the standard spin-dependent response is $\Sigma' + \Sigma''$. There are, however, two other important responses, as shown in Table~\ref{tab: responses}:  $\Delta$ and $\Phi''$.  These novel responses correspond to a coupling to the orbital angular momentum and to the orbital-spin interaction of the nucleus, respectively.  The sixth response, $\tilde{\Phi}'$, arising only in CP non-conserving interactions, does not appear in any of the models we consider.  It is difficult to find a UV model in which this response last response arises \cite{Fitzpatrick:2012ix}.

\begin{table}
\begin{tabular}{l l >{$}l<{$}}
$X$ & & {4 \pi \over 2 J + 1} W_X^{(p,p)}(0)\\
\hline
$M$~~~~~	& spin-independent & Z^2 \\
$\Sigma''$		& spin-dependent (longitudinal)  & 4 {J+1 \over 3 J} \langle S_p \rangle^2 \\
$\Sigma'$		& spin-dependent (transverse) 	 & 8 {J+1 \over 3 J} \langle S_p \rangle^2\\
$\Delta$		& angular-momentum-dependent & {1 \over 2} {J+1 \over 3 J} \langle L_p \rangle^2 \\
$\Phi''$		& angular-momentum-and-spin-dependent  & \sim \langle \vec{S}_p \cdot \vec{L}_p \rangle^2 \footnote{See Table 1 of \cite{Fitzpatrick:2012ix}.}\\ 
\hline
\end{tabular}
\caption{ Summary of the five nuclear responses relevant for DM direct detection. We also include the $q^2\rightarrow0$ limit of the associated response function, ${4 \pi \over 2 J + 1} W_X^{(N,N')}$, for $N=N'=p$. The response functions $W$ are as defined in Eq.~41 of \cite{Anand:2013yka}. Responses $M$ and $\Phi''$ can interfere, as can $\Sigma'$ and $\Delta$. In the $q^2\rightarrow0$ limit, ${4 \pi \over 2 J + 1}W_{\Delta \Sigma'}^{(N,N')} \rightarrow -2 {J+1 \over 3 J}\langle L_N\rangle \langle S_{N'}\rangle$. The response entering into ``standard'' spin-independent scattering is $M$ while that entering into ``standard'' spin-dependent scattering is $\Sigma'' + \Sigma'$. As in, \cite{Fitzpatrick:2012ix}, we will refer to $\Delta$ and $\Phi''$ as ``novel'' responses. }\label{tab: responses}
\end{table}

The purpose of the present paper is  to assess the impact of the new nuclear responses on scattering rates by examining a set of benchmark models motivated by relativistic operators that can be easily UV completed.  
We consider the relativistic operators summarized in Table \ref{tab: rel ops} along with their non-relativistic reductions and dependence on nuclear responses. We consider anapole, magnetic dipole, and electric dipole interactions, with coupling to the electromagnetic (EM) current arising due to {\em e.g.} kinetic mixing of a dark gauge field with the Standard Model electromagnetic $U(1)$. The anapole is attractive because it is the leading operator through which Majorana DM can couple to the nucleus through a vector interaction.  The electric and magnetic dipoles couple the DM spin to the field strength, and naturally arise in some models of composite DM \cite{Banks:2010eh,Bagnasco:1993st}. We also consider momentum-dependent interactions that can arise {\em e.g.} if the DM-nucleon interaction is mediated by a pseudoscalar---perhaps a pseudo-Goldstone boson \cite{Chang:2009yt}. We also consider a model sketched in \cite{Fitzpatrick:2012ix}, for which the novel spin-and-angular-momentum-dependent response, $\Phi''$, is important. A complete catalog of relativistic operators relevant for scattering, along with their non-relativistic reductions can be found in \cite{Fitzpatrick:2012ix} and \cite{Anand:2013yka}. See also \cite{DelNobile:2013sia}.

\begin{table}
\begin{tabular}{c >{$}r<{$} @{=} >{$}l<{$} >{$}c<{$} >{$}c<{$}}
\hline
~~Model~~ & \multicolumn{2}{c}{Relativistic Ops.} 																& \text{Nonrel.~Ops.} 	& \text{Resp.}	\\
\hline 
\multirow{2}{*}{pseudo-}& \oprel_2			& i\chibar \chi \Nbar \gamma^5 N								& \op_{10}	 =i\vec{S}_N\cdot{\vec{q} \over m_N}			&  \Sigma''			\\ 
\multirow{2}{*}{mediated}& \oprel_3			& i\chibar \gamma^5 \chi \Nbar N									& \op_{11}	=i \vec{S}_\chi\cdot {\vec{q} \over m_N}			&  M				\\
& \oprel_4			&  \chibar \gamma^5 \chi \Nbar \gamma^5 N													& \op_6=(\vec{S}_\chi \cdot {\vec{q} \over m_N})(\vec{S}_N \cdot {\vec{q} \over m_N}) 					&  \Sigma''			\\
\hline 
magnetic & \op_{9}^\text{rel}	&  \chibar i \sigma^{\mu \nu} {q_\nu \over m_M} \chi {K_\mu \over m_M} \bar{N} N 						& \op_1=\mathbf{1}_\chi \mathbf{1}_N , \op_5=i\vec{S}_\chi \cdot({\vec{q} \over m_N} \times \vec{v}^\perp) 		& M, \Delta		\\
dipole  &\op_{10}^\text{rel}	& \chibar i \sigma^{\mu \nu} {q_\nu \over m_M} \chi \bar{N} i \sigma_{\mu \alpha} {q^\alpha \over m_M} N 	& \op_4 =\vec{S}_\chi \cdot \vec{S}_N	, \op_6		& \Sigma'', \Sigma'	\\ 
 \hline
\multirow{2}{*}{anapole}& \op_{13}^\text{rel}	& \chibar \gamma^\mu \gamma^5 \chi {K_\mu \over m_M} \Nbar N									& \op_8=\vec{S}_\chi \cdot \vec{v}^\perp			& M, \Delta		\\  
& \op_{14}^\text{rel}	& \chibar \gamma^\mu \gamma^5 \chi \Nbar {i \sigma_{\mu \nu} q^\nu \over m_M} N					& \op_9=i \vec{S}_\chi \cdot(\vec{S}_N \times {\vec{q} \over m_N})			& \Sigma'			\\ 
\hline
electric & \op_{17}^\text{rel}	& i{P^\mu \over m_M}\chibar \gamma^\mu \gamma^5 \chi {K_\mu \over m_M} \Nbar N			& \op_{11}	=i \vec{S}_\chi\cdot {\vec{q} \over m_N}		& M				\\
dipole & \op_{18}^\text{rel}	& i{P^\mu \over m_M}\chibar \gamma^\mu \gamma^5 \chi \Nbar {i \sigma_{\mu \nu} q^\nu \over m_M} N	& \op_{11},\op_{15}=-\left( \vec{S}_\chi \cdot {\vec{q} \over m_N}\right)\left(( \vec{S}_N \times \vec{v}^\perp) \cdot {\vec{q} \over m_N}\right) 	& M,\Phi'',\Sigma'			\\
\hline
\multirow{2}{*}{$\vec{L}\cdot\vec{S}$-} & \op_{5}^\text{rel}	& {P^\mu \over m_M}\chibar \chi {K_\mu \over m_M} \Nbar N			& \op_{1}		& M				\\
\multirow{2}{*}{generating} & \op_{6}^\text{rel}	& {P^\mu \over m_M}\chibar \chi \Nbar {i \sigma_{\mu \nu} q^\nu \over m_M} N	& \op_{1},\op_{3}=i \vec{S}_N \cdot \left( {\vec{q} \over m_N} \times \vec{v}^\perp\right) 	& M,\Phi'',\Sigma'			 \\
 &  \multicolumn{2}{c}{and $\op_{10}^\text{rel}$ (see above)} 		\\
\hline
\end{tabular}
\caption{Relativistic operators from Table 1 of \cite{Anand:2013yka} (v1) that we consider in this work, grouped according to the linear combinations that we consider together. Here $K=k+k'$ where $k$ and $k'$ are the incoming and outgoing four-momenta of the nucleon $N$, respectively, and similarly for the DM momentum $P=p+p'$,  and $q$ is the four-momentum transfer ($q=k-k'=p'-p$).  
We also include the non-relativistic operators that appear in the non-relativistic reduction of the given relativistic operator. Note that $\op_1$ is the standard spin-independent operator and $\op_4$ is the standard spin-dependent operator. Finally, we also include the dependence on the five nuclear responses relevant for DM scattering, which are summarized in Table \ref{tab: responses}. See also \S \ref{approach to models} for discussion.
}\label{tab: rel ops}
\end{table}

The models we consider, besides being well motivated by UV completions, also encompass the most interesting operators in terms of probing the new nuclear responses.  As we will see explicitly below, different nuclei can have very different sensitivity to these new responses.  This can already be seen in the earlier 
work  of \cite{Fitzpatrick:2010br}, which utilized operators in a relativistic effective field theory.  The anapole interaction, for example, leads to a proton-orbital-angular-momentum response ($\Delta$), which, because of the stronger $\Delta$ response of sodium than germanium and xenon (see Table~\ref{tab: numerical responses}), can bring the DAMA region of interest into agreement with the CoGeNT region of interest, and simultaneously reduce the tension between DAMA and xenon-target experiments.  In the treatment of \cite{Fitzpatrick:2010br}, the stronger response of sodium is apparent simply because of its large nuclear magnetic moment.\footnote{The magnetic response is a particular combination of orbital angular momentum and spin responses.} The new responses, as the momentum transfer drops to zero, also only depend on the spin and orbital angular momentum of the nucleus, so that the new responses in this limit well reproduce the result in \cite{Fitzpatrick:2010br}, which neglects possible nonstandard momentum dependence of the nuclear response.  As the momentum transfer becomes large compared to inverse nuclear size, this kind of treatment breaks down.

Thus, while this ``standard treatment'' using operators in a relativistic effective field theory 
can work well in the low momentum transfer limit, 
the nuclear responses of \cite{Fitzpatrick:2012ix} must be employed at larger momentum transfer to correctly model the DM-nucleus interaction. Thus direct detection rates for weak scale or heavier DM, for which larger momentum transfer is relevant, can be more affected by the new nuclear responses than for low-mass DM, where the effect of the momentum dependence of the new responses is negligible.  

In addition, while the new nuclear responses of \cite{Fitzpatrick:2012ix} should correctly reproduce macroscopic properties of the nucleus like its spin and magnetic moment in the momentum transfer $q^2 \rightarrow 0$ limit, in practice the responses for some nuclei differ substantially from the measured result.  Thus comparing the the nuclear responses from \cite{Fitzpatrick:2012ix} against the treatment using operators in a relativistic effective field theory in the $q^2 \rightarrow 0$ limit can give one a good sense of the uncertainty in the nuclear responses computed in \cite{Fitzpatrick:2012ix}. In \S\ref{sec: response functions} we highlight the importance of this overall normalization of the nuclear responses by comparing the standard treatment against one incorporating the nuclear response functions of \cite{Fitzpatrick:2012ix} for the case of light DM.  Having separated out the uncertainty in the overall normalization of the nuclear response using light DM, we then concentrate on the importance of the momentum dependence of form factors in the context of heavy DM scattering rates.

Before moving into the general discussion, we highlight some practical points relevant for implementing nuclear responses in the context of DM direct detection. 
\begin{itemize}
\item The new nuclear responses are important for mediators that couple DM or nucleon spin to momentum transfer and/or velocity. 
\item Nuclei with unpaired nucleon spins have the most potential to give rise to different results than the standard treatment.  The usual spin-dependent form factors are a particular linear  combination of a larger set of independent operators and therefore do not represent the full range of nuclear responses.  
\item For low-mass DM (we consider the case of DM with mass $\sim$10 GeV), a standard treatment ignoring the momentum dependence of novel form factors is sufficient, and will reproduce the results with the full form factors. 
\item The most substantial differences due to momentum dependence of form factors arise for heavy elements such as iodine and xenon with abundant isotopes that have an unpaired nucleon, in momentum-/velocity-dependent interactions such as anapole and dipole interactions.  For example COUPP, which has a large, spin-dependent iodine target and is sensitive to a large recoil energy range, will have its differential rates substantially modified by the new nuclear form factors. Xenon target experiments would be more sensitive to the previously-ignored momentum dependence if they probed higher energies than current experiments, which are sensitive in the approximate range $E_R=4-30$ keV. 
\item Smaller differences arise in smaller elements such as germanium, and for yet smaller elements like fluorine and sodium, where the momentum dependence of form factors is practically negligible over the recoil energy range relevant for direct detection.  That said, $^{73}$Ge has a huge total angular momentum ($J=9/2$) and a huge contribution from orbital angular momentum, meaning that even the abundance-weighted orbital-angular-momentum response of germanium can be substantial and---if a germanium-based experiment were to probe an order 100+ keV energy range---could be important.
\end{itemize}

The outline of this paper is as follows.  In Sec.~\ref{approach to models} we analytically map our
UV-complete benchmark models onto the nuclear response functions.  In Sec.~\ref{sec: response functions}, we then examine the impacts on rates for each of our 
models, comparing the results with the new form factors to the results one would obtain using the standard form factors.  We begin in Sec.~\ref{sec: light DM} with a discussion of the overall normalization of nuclear response functions in the context of light DM; we  draw constraints from experiments such as LUX, XENON100, PICASSO, and CDMS as well as a few light DM regions of interest, utilizing the new and old responses. We move on to discussing the momentum dependence of novel form factors in Sec.~\ref{sec: momentum dependence of form factors}. To further illustrate the effects of the nuclear responses, in Sec.~\ref{sec: fits and bounds} we simulate the effects of the novel nuclear responses on a purported signal and also draw constraints for our benchmark models over a 1 TeV DM mass range. We conclude in Sec.~\ref{sec: conclusions}.

\section{Mapping models of momentum-dependent dark matter to nuclear responses}\label{approach to models}

Direct detection bounds have been analyzed for many of these models previously, as in \cite{Pospelov:2000bq,Bagnasco:1993st,Sigurdson:2004zp,Chang:2009yt,Feldstein:2009tr,Fitzpatrick:2010br,Banks:2010eh,An:2010kc,Barger:2010gv,Dienes:2013xya}. Here we provide a systematic, updated analysis, including a proper treatment of nuclear responses. 

In order to incorporate the novel nuclear responses and to adopt the more ``model-independent'' language of operator analyses, we use the nuclear response functions and conventions of \cite{Anand:2013yka}. The scattering rate given an interaction written in terms of the non-relativistic operators in Table \ref{tab: rel ops} can be deduced from Eqs.~38-40 of \cite{Anand:2013yka}. Specifically, for scattering off of a target, $T$,\footnote{Throughout this paper we use $T$ to denote target and $N, N'$ for nucleon ($N=p~\text{or}~n$).}
\beq
\sigma_T \equiv {2 \mu_T^2 v^2 \over m_T}{d \sigma_T \over d{E_R}} ={ \mu_T^2 \over \pi}\langle |\mathcal{M}|^2\rangle_\text{non-rel}^\text{Nuc}
\label{eq:crosssection}
\eeq
where \cite{Anand:2013yka} 
\begin{multline}
\langle |\mathcal{M}|^2\rangle_\text{non-rel}^\text{Nuc} =  \sum_{N,N'=p,n} \bigg[ \sum_{k=M,\Sigma',\Sigma''} R_k\left(v_T^{\perp 2}, {\vec{q}^{\,2} \over m_N^2}, c_i^{(N)}, c_j^{(N')}\right) \W_k^{(N,N')}(y)\\ +{\vec{q}^{\,2} \over m_N^2} \sum_{k=\Phi'', \Delta, M \Phi'', \Delta \Sigma'} R_k\left(v_T^{\perp 2}, {\vec{q}^{\,2} \over m_N^2}, c_i^{(N)}, c_j^{(N')}\right) \W_k^{(N,N')}(y) \bigg],
\label{eq:matrixelement}
\end{multline}
with $R_k$  encoding the momentum and velocity dependence coming from the WIMP interaction with the mediator of the interaction (depending on the non-relativistic operator coefficients $c_i^N$ as well as momentum transfer and WIMP velocity), and $\W_k$ the nuclear response functions depending on
\beq 
y=(q \, b/2)^2 = m_T E_R b^2 / 2, \qquad \text{where} \qquad b=\sqrt{41.467/(45 A^{-1/3}-25 A^{-2/3})} \; \text{fm}
\eeq
is nuclear size (following~\cite{Anand:2013yka}) for a target of atomic mass number $A$.
 Here, $\vec{v}_T^\perp = \vec{v}_T+\vec{q}/2 \mu_T$ where $\vec{v}_T$ is the DM velocity in the lab frame, and $\vec{v}_T^\perp$ has been defined so that $\vec{v}_T^\perp \cdot \vec{q} = 0$ and thus $\vec{v}_T^{\perp 2} = \vec{v}_T^2 - \vec{q}^{\,2}/4 \mu_T^2$.  For convenience we have defined 
\beq
\W_k={4 \pi \over 2 J + 1} W_k
\eeq where $J$ is nuclear spin, since the response functions $W$ as defined in \cite{Anand:2013yka} are always accompanied by the quantity ${4 \pi \over 2 J + 1}$. Expressions for the functions $R_k$, which link the nucleon-DM scattering coefficients to the nuclear responses, are provided in \eqref{eq: response coefficients}.

In general, the rate at which DM scatters off a given target, per target mass per recoil energy is given by,
\beq
{dR \over dE_R} = {1 \over m_T}{\rho_\chi \over m_\chi} \int_{v_\text{min}} v f(\vec{v}) {d\sigma_T \over dE_R} d^3\vec{v} \label{eq: rate}
\eeq
where $m_T$ is the target nucleus mass, $\rho_\chi$ is the local DM density, $m_\chi$ is the DM mass, and $f(\vec{v})$ is the local DM velocity distribution.  In subsequent sections, where relevant we will assume a Standard Halo Model (SHM) velocity distribution and density, with $
v_0=220~\text{km/s}$ and  
$v_\text{esc} = 544~\text{km/s}$.

Next we provide explicit formulae for scattering rates for the models we consider, utilizing the nuclear responses of \cite{Anand:2013yka}.

\subsection{Review of the standard spin-independent and -dependent cases.}

The standard spin-independent, WIMP-nucleus interaction can result from the effective Lagrangian,
\beq
\lagint^\text{SI} = \sum_{N=n,p} {f_\text{SI}^N \over \Lambda^2} \bar{\chi} \chi \bar{N} N \rightarrow \sum_{N=n,p} c_1^N \op_1 ~~\text{with}~~c_1^N={f_\text{SI}^N \over \Lambda^2} \label{std Lint},
\eeq leading to the cross section,
\beq
\sigma_T^\text{SI} = {\mu_T^2 \over \pi } {1 \over \Lambda^4}\left( {f_\text{SI}^p}^2 \W_M^{(p,p)} + 2 f_\text{SI}^p f_\text{SI}^n \W_M^{(p,n)} + {f_\text{SI}^n}^2 \W_M^{(n,n)} \right) \label{eq: SI rate}
\eeq
which is often expressed,
\beq
\sigma_T^\text{SI} = {\mu_T^2 \over \mu_p^2} \sigma_p^\text{SI} \left(Z + (A-Z)f_\text{SI}^n/f_\text{SI}^p  \right)^2 F^2 
\eeq
where the form factor $F^2$ is defined to be 1 at zero momentum transfer,
\beq
F^2(y) \equiv { {f_\text{SI}^p}^2 \W_M^{(p,p)}(y) + 2 f_\text{SI}^p f_\text{SI}^n \W_M^{(p,n)}(y) + {f_\text{SI}^n}^2 \W_M^{(n,n)}(y)  \over {f_\text{SI}^p}^2 \W_M^{(p,p)}(0) + 2 f_\text{SI}^p f_\text{SI}^n \W_M^{(p,n)}(0) + {f_\text{SI}^n}^2 \W_M^{(n,n)}(0)}
\eeq
and where $\sigma_p^\text{SI}$ is the zero-momentum-transfer cross section off of protons,\footnote{Note this is for non-Majorana dark matter; multiply by a factor of 4 for the Majorana case.} 
\beq\label{sigma p}
\sigma_p^\text{SI} = {\mu_p^2 \over \pi} \left(f_\text{SI}^p \over \Lambda^2\right)^2. 
\eeq

Standard spin-dependent (SD) scattering results from the effective lagrangian,
\beq
\lagint^\text{SD} = \chibar \gamma^\mu \gamma^5 \chi \sum_{N=n,p} {f_\text{SD}^N \over \Lambda^2} \Nbar \gamma^\mu \gamma^5 N \rightarrow \sum_{N=n,p} c_4^N \op_4 ~~\text{with}~~c_4^N=-{4 f_\text{SD}^N \over \Lambda^2} ,
\eeq
leading to the cross section,
\beq
\sigma_T^\text{SD}={\mu_T^2 \over \pi}{C_\chi \over \Lambda^4}\sum_{N,N'} f_\text{SD}^N f_\text{SD}^{N'} \left( \W_{\Sigma'}^{(N,N')} + \W_{\Sigma''}^{(N,N')} \right)
\eeq
where, as in \cite{Fitzpatrick:2012ix}, we have defined the DM spin-dependent constant,
\beq
C_\chi \equiv {4 j_\chi (j_\chi + 1) \over 3}.
\eeq
The SD cross section is often expressed as a function of the proton-DM zero-momentum-transfer cross section $\sigma_p^\text{SD}$,
\beq
\sigma_T^\text{SD}={\mu_T^2 \over \mu_p^2} \sigma_p^\text{SD} {4 \over 3}{J+1 \over J}\left( \langle S_p \rangle +  {f_\text{SD}^n \over f_\text{SD}^p}\langle S_n \rangle \right)^2 {\sum_{N,N'} f_\text{SD}^N f_\text{SD}^{N'} \left( \W_{\Sigma'}^{(N,N')}(y) + \W_{\Sigma''}^{(N,N')}(y) \right) \over \sum_{N,N'} f_\text{SD}^N f_\text{SD}^{N'} \left( \W_{\Sigma'}^{(N,N')}(0) + \W_{\Sigma''}^{(N,N')}(0) \right) },
\eeq
where  \beq \sigma_p^\text{SD}= {\mu_p^2 \over \pi} {C_\chi \over \Lambda^4} 3 (f_\text{SD}^p)^2. \eeq
Note that the combination of nuclear responses, 
\beq
\sum_{N,N'} f_\text{SD}^N f_\text{SD}^{N'} \left( \W_{\Sigma'}^{(N,N')}(0) + \W_{\Sigma''}^{(N,N')}(0) \right) = 4 {J + 1 \over J} (f_\text{SD}^p \langle S_p \rangle + f_\text{SD}^n \langle S_n \rangle)^2,
\eeq
gives rise to the usual spin-dependent factors.


\subsection{Anapole Dark Matter}

Majorana fermion DM scattering off of nucleons via a spin-1 mediator that kinetically mixes with the photon proceeds via the following effective interaction:\footnote{The non-relativistic reduction for this and other interactions considered in the paper can be read from Table 1 of \cite{Anand:2013yka}. To do so, one must recall the Gordon identities, $\bar{u}(p') \gamma^\mu u(p) = \bar{u}(p') \left({(p+ {p'})^\mu \over 2 m} + {i \sigma^{\mu \nu} (p'-p)_\nu \over 2 m} \right) u(p)$ and $\bar{u}(p') \sigma^{\mu \nu} (p'-p)_\nu \gamma^5 u(p) =  \bar{u}(p') \left( { i(p+ {p'})^\mu} \gamma^5  \right) u(p)$. Note that signs in Table 1 in v1 of \cite{Anand:2013yka} for the non-relativistic reduction of relativistic operators with an odd power of momentum transfer are incorrect by a factor of -1, because the convention $q = p-p'$ was used rather than the stated $q=p'-p$ convention.}
\begin{equation}
\lagint^\text{anapole} = {f_a \over  M^2} \chibar \gamma^\mu \gamma^5 \chi \mathcal{J}^\text{EM}_\mu\\
\end{equation}
where 
\begin{equation}
\mathcal{J}^\text{EM}_\mu \equiv \sum_{N=n,p} \bar{N} \left(Q_N {K_\mu \over 2 m_N} - \mag_N {i \sigma_{\mu \nu} q^\nu \over 2 m_N} \right)N
\end{equation}
is the electromagnetic current restricted to nucleons. We have used the conventions of \cite{Anand:2013yka}, taking $K^\mu = k^\mu + {k'}^\mu$ and four-momentum-transfer $q^\mu = {p'}^\mu - p^\mu = k^\mu - {k'}^\mu$ with $p$($p'$) the incoming(outgoing) DM four-momentum and similarly $k$($k'$) the incoming(outgoing) nucleon four-momentum. We have used $\mag$ to denote a dimensionless magnetic moment,
\beq
\mag = { \text{magnetic moment} \over \text{nuclear magneton}} .
\eeq
The relevant EM constants are $\mag_n=-1.9$, $\mag_p=2.8$, $Q_p=1$, and $Q_n=0$.

In the non-relativistic limit, 
\begin{equation}
\lagint^\text{anapole} \rightarrow {2 f_a \over  M^2} \sum_{N=n,p} \left( Q_N \op_8 + \mag_N \op_9 \right)\\ \label{anapole coeffs}
\end{equation} where the non-relativistic operators $\op_8$ and $\op_9$ are as defined in \cite{Anand:2013yka} and Table~\ref{tab: rel ops}. 

Evaluating Eq.~\ref{eq:matrixelement}, taking $c_8,~c_9$ from Eq.~\ref{anapole coeffs}, and substituting the ``WIMP form factors'' $R_k$ found in \cite{Anand:2013yka} and reproduced in Appendix \ref{sec: response coefficients}, we obtain (for Dirac DM) 
\begin{multline}
\sigma_T^\text{anapole}  = {{\mu_T^2} \over \pi}\left({f_a \over M^2}\right)^2 C_\chi \bigg\{ \vec{v}^{\perp 2}_T \W_M^{(p,p)} + \\
{\vec{q}^{\,2} \over m_N^2} \left[\W_\Delta^{(p,p)}- \mag_n \W_{\Delta \Sigma'}^{(p,n)}-\mag_p \W_{\Delta\Sigma'}^{(p,p)} + {1 \over 4} \left( \mag_p^2 \W_{\Sigma'}^{(p,p)}+2 \mag_n \mag_p \W_{\Sigma'}^{(p,n)}+\mag_n^2 \W_{\Sigma'}^{(n,n)}\right) \right] \bigg\}
\label{eq: anapole rate}  
\end{multline}
where $C_\chi \equiv 4 j_\chi (j_\chi+1)/3 $.  The shell model predicts that the magnetic moment of a nucleus, $T$, is given by 
\beq
\mag_T = 2 \mag_p \langle S_p \rangle+2 \mag_n \langle S_n \rangle+ \langle L_p \rangle. \label{eq: mag moment}
\eeq
Referring to Table \ref{tab: responses}, one can check that in the $q^2 \rightarrow 0$ limit, the term in square brackets goes to ${J+1 \over 6 J} \mag_T^2$ and $\W_M^{(p,p)} \rightarrow Z^2$.  In this limit, Eq.~\ref{eq: anapole rate} reproduces the cross-section derived in \cite{Fitzpatrick:2010br}:
\begin{equation}
\sigma_T^\text{anapole}  = {{\mu_T^2} \over \pi}\left({f_a \over M^2}\right)^2 \left((\vec{v}^{\,2}-\frac{\vec{q}^{\,2}}{4 \mu_T^2}) Z^2 F(E_R)^2 + \vec{q}^{\,2} \frac{J+1}{6 J} \frac{\tilde{\mu}_T^2}{m_N^2}  \right).  
\end{equation}
When drawing bounds or regions of interest, we will parameterize the anapole coupling strength via $\tilde{\sigma} = f_a^2 \mu_p^2/\pi M^4$.

\subsection{Dipole-Interacting Dark Matter}
We next consider Dirac fermion DM that acquires dipole moments so that the effective WIMP-nucleon interaction is given by
\begin{align} \label{magdip lint}
\lagint^\text{magnetic dipole}&={f_\text{md} \over M^2} \chibar {i \sigma^{\mu \nu}  q_\nu \over \Lambda} \chi \mathcal{J}^\text{EM}_\mu \\
&\rightarrow {2 f_\text{md} \over M^2}\sum_{N=n,p} \left( Q_N \left( {m_N \over \Lambda} \op_5 - {\vec{q}^{\,2} \over 4 m_\chi \Lambda} \op_1 \right) + \mag_N \left( {m_N \over \Lambda} \op_6 - {\vec{q}^{\,2} \over m_N \Lambda} \op_4 \right) \right). \label{eq:c1c4}
\end{align}

Here again, we evaluate Eq.~\ref{eq:matrixelement}, taking $c_1,~c_4,~c_5,~c_6$ from Eq.~\ref{eq:c1c4}, and substitute the ``WIMP form factors'' $R_k$ of \cite{Anand:2013yka} to obtain 
\begin{multline}
\sigma_T^\text{magnetic dipole}  = {{\mu_T^2} \over \pi}\left({f_\text{md} \over M^2}\right)^2 {\vec{q}^{\,2} \over \Lambda^2} \bigg\{ \left[ C_\chi \vec{v}^{\perp 2}_T + {\vec{q}^{\,2} \over 4 m_\chi^2} \right]\W_M^{(p,p)} + \\
C_\chi {\vec{q}^{\,2} \over m_N^2} \left[\W_\Delta^{(p,p)}- \mag_n \W_{\Delta \Sigma'}^{(p,n)}-\mag_p \W_{\Delta\Sigma'}^{(p,p)} + {1 \over 4} \left( \mag_p^2 \W_{\Sigma'}^{(p,p)}+2 \mag_n \mag_p \W_{\Sigma'}^{(p,n)}+\mag_n^2 \W_{\Sigma'}^{(n,n)}\right) \right] \bigg\}.
\label{eq: mag dip rate}   
\end{multline}
As for Eq.~\ref{eq: anapole rate}, one can verify that in the $q^2 \rightarrow 0$ limit, we reproduce the results of \cite{Fitzpatrick:2010br}:\footnote{Up to a of 4 factor having to do with the normalization of operator coefficients.}
\begin{equation}
\sigma_T^\text{magnetic dipole}  = {{\mu_T^2} \over \pi}\left({f_{\rm md} \over M^2}\right)^2 {\vec{q}^{\,2} \over \Lambda^2} \left(\left(\vec{v}^{\,2}-{\vec{q}^{\,2} \over 4} \left(\frac{2}{m_T m_\chi}+\frac{1}{m_T^2}\right)\right) Z^2 F(E_R)^2 + \vec{q}^{\,2} \frac{J+1}{6 J} \frac{\tilde{\mu}_T^2}{m_N^2}  \right). 
\end{equation}
As in \cite{Gresham:2013mua}, when drawing bounds or regions of interest, we will parameterize the magnetic dipole coupling strength via $\tilde{\sigma} = f_\text{md}^2 \mu_p^2/\pi M^4$ and take $\Lambda=1$ GeV.

Likewise, the electric dipole reduces to,
\begin{align} \label{elecdip lint}
\lagint^\text{electric dipole}&={f_\text{ed} \over M^2} \chibar {\sigma^{\mu \nu}  q_\nu \gamma^5 \over \Lambda} \chi \mathcal{J}^\text{EM}_\mu\\
&\rightarrow {2 f_\text{ed} \over M^2}\sum_{N=n,p} \left( -Q_N  {m_N \over \Lambda} \op_{11} + \mag_N \left( {m_N \over \Lambda} \op_{15} + {m_\chi \vec{q}^{\,2} \over 4 m_N^2 \Lambda} \op_{11} \right) \right). 
\end{align} 
Similarly to the anapole and magnetic dipole, this reduces to 
\begin{equation}
\sigma_T^\text{electric dipole}  = {{\mu_T^2} \over \pi}\left({f_\text{ed} \over M^2}\right)^2 {\vec{q}^{\,2} \over \Lambda^2}{C_\chi} \left(  \W_M^{(p,p)} + \text{terms of order }{\vec{q}^{\,2} \over m_N^2} \right).
 \label{hax elec dip nuc rate}  
\end{equation}
For the electric dipole, the interesting terms depending on the novel response function $W_{\Phi''}$ (arising from $\op_{15}$) are momentum-suppressed compared to the spin-independent term. Thus, at the low energies relevant for direct detection, the cross section has the same form as the momentum-suppressed, spin-independent  ``pseudoscalar-mediated'' cross section considered in \cite{Chang:2009yt} and below, and is an example of how a momentum-suppressed spin-independent interaction could naturally arise with proton-only ``photonic''  \cite{Feng:2013vaa} couplings. In Sec.~\ref{sec: response functions}, we will use the momentum-suppressed, spin-independent case ($q^2\times\text{SI}$) to establish what sort of typical error to expect in form factors at larger momentum transfer by comparing results for the $q^2\times\text{SI}$ rate using either the spin-indepentent ($M$) form factors of  \cite{Fitzpatrick:2012ix,Anand:2013yka} or using the Helm form factor.

\subsection{$(\vec{L}\cdot \vec{S})$-Generating}

In the case of both the anapole and magnetic dipole operators, the new response $\Delta$, as well as $\Sigma'$ (which is not the usual spin-dependent combination $\Sigma'+\Sigma''$), compete with, and in some cases dominate over, the charge form factor $M$.  By contrast, the new $\Phi''$ response in the electric dipole operator is suppressed by $q^2/m_N^2$ in comparison to the charge form factor, so that, unless the mediator couples only to the neutron, the standard form factor $M$ always dominates in the electric dipole operator.  Here we consider what types of interactions allow the $(\vec{L}\cdot \vec{S})$-Generating $\Phi''$ response to dominate, when the contribution from $M$ is subdominant.  In particular, we consider the interaction highlighted in \cite{Fitzpatrick:2012ix},
\begin{align}
\lagint^\text{LS} &= {f_\text{LS} \over \Lambda^2} \bar{\chi} \gamma_\mu \chi \sum_{N=n,p}\left( \kappa_1^N {q_\alpha q^\alpha \over m_N^2} \Nbar \gamma^\mu N + \kappa_2^N \Nbar {i \sigma^{\mu \nu} q_\nu \over 2 m_N} N \right)\\
&\rightarrow {f_\text{LS} \over \Lambda^2} \sum_{N=n,p} \left( \left( {\kappa_2^N \over 4} - \kappa_1^N\right){\vec{q}^{\,2} \over m_N^2} \op_1 - \kappa_2^N \op_3 + \kappa_2^N {m_N \over m_\chi} \left( {\vec{q}^{\,2} \over m_N^2} \op_4 - \op_6\right)\right).
\end{align}

From Eqs.~38-40 of \cite{Anand:2013yka} 
\begin{multline}
\sigma_T^\text{LS}  = {{\mu_T^2} \over \pi}\left({f_\text{LS} \over \Lambda^2}\right)^2 {\vec{q}^{\,2} \over m_N^2} \sum_{N,N'} \Bigg( {\vec{q}^{\,2} \over m_N^2} \bigg\{ \left(\kappa_1^N-{\kappa_2^N \over 4}\right)\left(\kappa_1^{N'}-{\kappa_2^{N'} \over 4}\right)\W_M^{(N,N')} + \\
\kappa_2^N\left(  \kappa_1^{N'}-{\kappa_2^{N'} \over 4}\right) \W_{\Phi'' M}^{(N,N')} + {\kappa_2^N \kappa_2^{N'} \over 4} \left[ \W_{\Phi''}^{(N,N')} + {C_\chi \over 4} {m_N^2 \over m_\chi^2} \W_{\Sigma'}^{(N,N')} \right]  \bigg\} + {\vec{v}_T^{\perp \,2}} {\kappa_2^N \kappa_2^{N'} \over 8}\W_{\Sigma'}^{(N,N')} \Bigg). 
\label{eq: LS rate}  
\end{multline}
To parameterize the overall coupling strength we will use $\tilde{\sigma} = f_\text{LS}^2 \mu_p^2/\pi \Lambda^4$. We will consider the case where 
\beq
\kappa_1^N-{\kappa_2^N \over 4} = 0 \qquad \text{and} \qquad \kappa_2^p = \kappa_2^n=2. \label{eq: LS params}
\eeq
Of the target elements we examine in this paper, for all but fluorine the $\Phi''$ response dominates over the $\Sigma'$ response (see Table~\ref{tab: numerical responses}). Even for fluorine the ${\vec{v}_T^{\perp \,2}}$ term becomes negligible for recoil energies of order 1 keV and above. Therefore we compute rates without including the ${\vec{v}_T^{\perp \,2}}$ term.

\subsection{Pseudoscalar-Mediated Dark Matter} \label{sec: pseudoscalar cross-sections}

\beq \label{pseudo lint}
\lagint^\text{pseudoscalar}={1 \over M^2} \sum_{N=n,p} \left( f_1^N i \chibar \gamma^5 \chi  \Nbar N + f_2^N i \chibar \chi  \Nbar \gamma^5 N+ f_3^N \chibar \gamma^5 \chi  \Nbar \gamma^5 N \right)
\eeq
The terms in \eqref{pseudo lint} are included in decreasing order of importance: if $f_1$, $f_2$, and $f_3$ are comparable, the $f_1$ term dominates over the $f_2$ term, which dominates over the $f_3$ term. This is because the $f_1$ term leads to a ${q}^2$-suppressed spin-independent interaction, the $f_2$ term to a ${q}^2$-suppressed spin-dependent interaction, and the $f_3$ term to a ${q}^4$-suppressed spin-dependent interaction. We thus consider each 
term separately, and focus on the isospin benchmark: $f_i^n=f_i^p$. If the DM is a scalar, only the $f_2$ term survives, and an overall factor of $m_\chi/2$ in comparison to the fermionic case enters into the matrix element. 

The non-relativistic reductions of the relevant operators are given by\footnote{See Table 1 of \cite{Anand:2013yka}.} 
\begin{align}
i \chibar \gamma^5 \chi  \Nbar N &\rightarrow -{m_N \over m_\chi} \op_{11}\\
i \chibar \chi  \Nbar \gamma^5 N &\rightarrow \op_{10} \\
\chibar \gamma^5 \chi  \Nbar \gamma^5 N &\rightarrow -{m_N \over m_\chi} \op_6,
\end{align}
and the associated cross section is,
\begin{multline}
\sigma_T^\text{pseudoscalar}  = {{\mu_T^2} \over \pi}\left({1 \over M^2}\right)^2\sum_{N,N'} \bigg( {\vec{q}^{\,2} \over 4 m_\chi^2} C_\chi  f_1^N f_1^{N'} \W_M^{(N,N')}  \\
 + \left[  {\vec{q}^{\,2} \over 4 m_N^2}f_2^N f_2^{N'} + {\vec{q}^{\,4} \over 16 m_N^2 m_\chi^2}C_\chi f_3^N f_3^{N'} \right]\W_{\Sigma''}^{(N,N')} \bigg) \label{eq: pseudoscalar rate}  
\end{multline}
where again $C_\chi \equiv 4 j_\chi (j_\chi+1)/3 $.

Note that the spin-dependent part of the interaction depends on only the longitudinal spin-dependent response rather than the longitudinal plus transverse spin-dependent response that is standardly considered for spin-dependent DM interactions. Thus for $y \sim 1$ it is inappropriate to treat the cross section arising from the $f_2$ or $f_3$ term as $(q^2 / q_\text{ref}^2)^n \times \sigma_T^\text{SD}$.

When drawing bounds or regions of interest, we will parameterize coupling strength via 
\beq
\tilde{\sigma} = {\mu_p^2 \over \pi} C_\chi {q_\text{ref}^2 \over 4 m_\chi^2} {(f_1^p)^2 \over M^4}, {\mu_p^2 \over \pi} {q_\text{ref}^2 \over 4 m_p^2} {(f_2^p)^2 \over M^4}, ~\text{or}~{\mu_p^2 \over \pi} C_\chi {q_\text{ref}^4 \over 16 m_p^2 m_\chi^2} {(f_3^p)^2 \over M^4}
\eeq with $q_\text{ref}=1$ GeV in the case where the $f_1$, $f_2$, or $f_3$ term dominates, respectively. Note that 
\beq
\sum_{N,N'} f^N f^{N'}\W_{\Sigma''}^{(N,N')}(0)={4 \over 3}{J+1 \over J}\left( f^p \langle S_p \rangle + f^n \langle S_n \rangle \right)^2.
\eeq


\section{Nuclear response functions and direct detection}\label{sec: response functions}

We now look at the concrete numerical impacts of the new nuclear responses on DM direct detection.  As discussed in the introduction, there are two factors to consider in evaluating their effects: the nuclear response at zero momentum transfer (which should be describable by macroscopic quantities such as the charge of the nucleus, as well as the proton and neutron spins) and the momentum dependence of the nuclear responses.  We separately evaluate these effects by separately considering the impacts of the nuclear response functions on light and heavy DM.  In both cases we consider both constraints and possible signals. 

\subsection{Nuclear responses at zero momentum transfer and light DM}\label{sec: light DM}

Lighter DM transfers less momentum to the nucleus in the interaction, and hence provides a good laboratory for studying the $q^2 \rightarrow 0$ limit of the nuclear responses.  Thus we begin by considering the regions of interest (ROIs), as highlighted by the DAMA and CoGeNT experiments, and constraints, as highlighted by the LUX and PICASSO experiments, for light DM.\footnote{Refs.~\cite{Fitzpatrick:2012ix,Anand:2013yka} do not include silicon form factors in their analysis and so we do include CDMS Silicon ROIs.} We will consider more massive DM in the next subsection.  
Bounds and ROIs for the aforementioned existing experiments are shown in Fig.~\ref{fig: light DM} using the procedure described in the appendix of \cite{Gresham:2013mua}. The solid lines show bounds/ ROIs exactly as calculated in \cite{Gresham:2013mua}, where for the expected differential rates only the Helm form factor was employed, where spin matrix elements were taken from the literature (\cite{Dimitrov:1994gc,Ressell:1997kx,Divari:2000dc,Menendez:2012tm}, summarized in \cite{Cannoni:2012jq}), and where measured magnetic moments\footnote{See {\em e.g.} WebElements.com or \cite{Ressell:1997kx}.} were employed in the anapole and magnetic dipole rates; no spin or angular momentum form factor momentum dependence was included (indicated by ``no SD form factors''). The thick dotted lines show bounds/ROIs derived using Eqs.~\eqref{eq: SI rate}, \eqref{eq: pseudoscalar rate}, \eqref{eq: anapole rate}, and \eqref{eq: mag dip rate} for the rates and employing the response functions $\W$ of \cite{Fitzpatrick:2012ix,Anand:2013yka}. The thin dashed lines are a sort of hybrid, employing the response functions of \cite{Fitzpatrick:2012ix,Anand:2013yka} but (re)normalized at $q^2\rightarrow0$ to match the $q^2\rightarrow0$ values of the rates in \cite{Gresham:2013mua}. The ``no SD form factors'' and ``renormalized form factors'' curves are essentially indistinguishable, which shows that the momentum dependence of the spin- and angular-momentum-dependent form factors is playing a negligible role for light DM. Furthermore, while there is no perceptible difference between the ``no SD form factors'' and ``full form factors'' curves in most cases, there are a few notable exceptions, which we discuss next. 

Given a $q^4$-suppressed longitudinal spin-dependent interaction, the full and no form factor results for DAMA (Na-I target) and PICASSO (F target) match very well, while those for LUX (Xe target) and CoGeNT (Ge target) do not. The primary reason that the results match for the Na and F targets and not for the Xe and Ge targets is that the $\langle S_{p(n)} \rangle$ values as implied by the $q^2 \rightarrow0$ limits of the response functions of \cite{Fitzpatrick:2012ix,Anand:2013yka} match the state-of-the-art values used in the ``no SD form factors'' calculation for F and Na but are quite different for Ge and Xe. This is unsurprising given that, according to \cite{Fitzpatrick:2012ix} (see also the discussion in \cite{Cannoni:2012jq}), the nuclear shell calculation used is much less sophisticated than the state-of-the-art for Ge, I, and Xe.  These differences can be taken as a gauge in the errors on the form factors themselves. In Table~\ref{tab: spins etc} we provide a summary of the theoretical spin and orbital angular momentum matrix elements implicit in the response functions of \cite{Fitzpatrick:2012ix,Anand:2013yka} (see Table~\ref{tab: responses} and Eq.~\ref{eq: mag moment}) and as given by the most advanced calculations that are reported in the literature. We also provide the theoretical and experimental values for magnetic moments.

Likewise, the mass-independent difference for PICASSO given magnetic dipole or anapole interactions derives from a 10\% difference, as shown in Table~\ref{tab: spins etc}, between the empirical magnetic moment and the ``theoretical'' magnetic moment implicit in the response functions we employ for fluorine.  On the other hand the measured and theoretical magnetic moments for sodium match almost exactly, explaining why the DAMA ROIs line up well.  The empirical and theoretical magnetic moments differ more substantially for relevant xenon and germanium isotopes, but since the anapole and magnetic dipole rates for scattering off of these targets are dominated by the DM-charge (spin-independent) interaction, it makes a negligible difference. 

\begin{figure}
\includegraphics{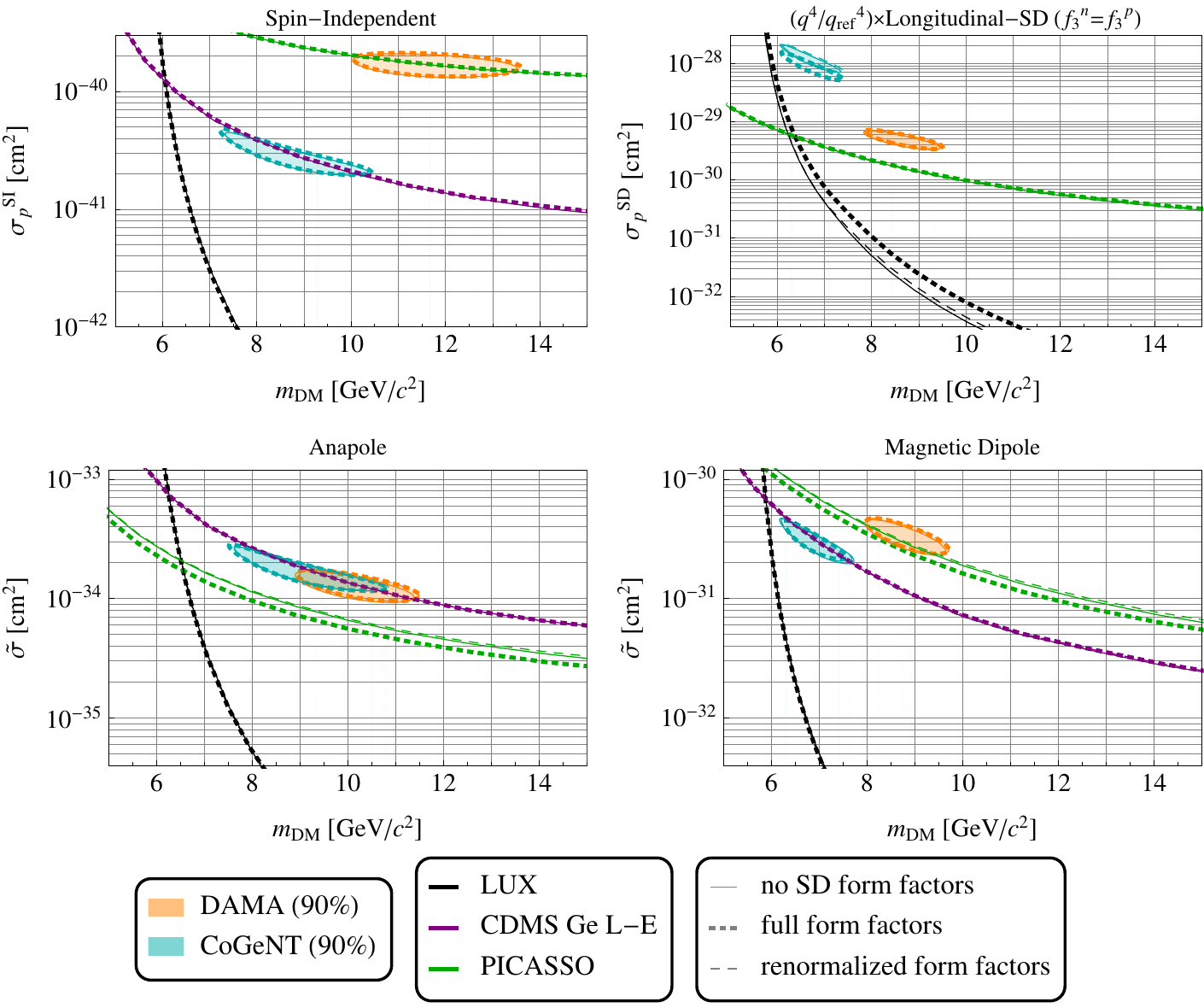}
\caption{ Limits and regions of interest for a representative set of direct detection experiments. Thick, dotted lines are derived using the full form factors provided in \cite{Anand:2013yka}, thin solid lines are those derived as described in \cite{Gresham:2013mua}, employing only the helm form factor as the charge-dependent form factor and no spin-dependent form factors, and thin dashed lines were derived using the ``full form factors'' of \cite{Anand:2013yka} but (re)normalized to the \cite{Gresham:2013mua} values at $q^2\rightarrow0$.}\label{fig: light DM}
\end{figure}

\subsection{Momentum dependence of nuclear responses and heavy DM}\label{sec: momentum dependence of form factors}

Above we have discussed uncertainty associated with the overall normalization of nuclear responses. Now we turn to discussing uncertainty associated with the momentum dependence of nuclear responses and, in particular, the relative importance of possibly novel momentum dependence encapsulated in novel form factors. 

The novel responses $\Delta$ and $\Phi''$ depend on different macroscopic properties of target nuclei than just spin or charge/mass number. To give an idea of which nuclei may be most sensitive to the new responses, in Table~\ref{tab: numerical responses} we show the value of the natural-abundance-weighted responses of fluorine, sodium, germanium, iodine, and xenon at zero momentum transfer, as calculated using the response functions of \cite{Anand:2013yka}. (Keep in mind, however, that the values especially for germanium, iodine, and xenon may be somewhat inaccurate due to limitations of the nuclear calculations performed for \cite{Fitzpatrick:2012ix}. See Table~\ref{tab: spins etc} and the discussion above and in \cite{Fitzpatrick:2012ix}.) The orbital-angular-momentum response $\Delta$ is particularly interesting because it deviates from the patterns of the spin-dependent and charge-dependent responses; the hierarchy of response strength of fluorine, sodium, germanium, iodine, and xenon for scattering off of protons or neutrons is \emph{quite different} than the hierarchy of response strengths for the standard spin-independent and -dependent responses. This is because the $\Delta$ response is sensitive also to the angular momentum of the orbital shell occupied by the unpaired nucleon \cite{Fitzpatrick:2012ix}, making $^{73}$Ge and $^{127}$I particularly sensitive in comparison to $^{19}$F and $^{129,131}$Xe, respectively. An interesting aspect of the $\Phi''$ response is that it relates to the occupation levels of orbitals and can be nonzero even for nuclei with zero total angular momentum \cite{Fitzpatrick:2012ix}, though for the target nuclei highlighted here the hierarchy of $\Phi''$ response strengths approximately tracks the relative strengths of the standard SI response---with a notable exception being the stronger response of iodine than xenon for scattering off of protons.  Indeed iodine is particularly sensitive to both novel nuclear responses.

\begin{table}
\begin{tabular}{c c c |  >{$}c<{$}  >{$}c<{$}  | >{$}c<{$}  >{$}c<{$} | c c c | c }
		& NA(\%)	& $J$	& \begin{matrix} |\sp_\text{th}| \\ |\sn_\text{th}| \end{matrix} &  \begin{matrix} \sp_\text{lit} \\ \sn_\text{lit} \end{matrix}~~~& \begin{matrix} |\Lp_\text{th}| \\ |\Ln_\text{th}| \end{matrix} &  \begin{matrix} \Lp_\text{lit} \\ \Ln_\text{lit} \end{matrix} & $|\tilde{\mu}_\text{th}|$ & $\tilde{\mu}_\text{lit}$ & $\tilde{\mu}_\text{exp}$ & lit Ref. \\
		\hline \hline
$^{19}$F & 100 & 1/2 & \begin{matrix}0.475\\0.009\end{matrix} & \begin{matrix}0.4751\\-0.0087\end{matrix} & \begin{matrix}0.224\\0.19\end{matrix} & \begin{matrix}0.4751\\-0.0087\end{matrix} & 2.911 & 2.91 & 2.6289 & \cite{Divari:2000dc} \\ \hline
$^{23}$Na & 100 & 3/2 & \begin{matrix}0.248\\0.02\end{matrix} & \begin{matrix}0.2477\\0.0199\end{matrix} & \begin{matrix}0.912\\0.321\end{matrix} & \begin{matrix}0.2477\\0.0199\end{matrix} & 2.219 & 2.22 & 2.2175  & \cite{Divari:2000dc}\\  \hline
$^{73}$Ge & 7.7 & 9/2 & \begin{matrix}0.008\\0.475\end{matrix} & \begin{matrix}0.03\\0.378\end{matrix}  & \begin{matrix}0.184\\3.832\end{matrix} & \begin{matrix}0.361\\3.732\end{matrix} & 1.591 & -0.92 & -0.8795 & \cite{Dimitrov:1994gc}\\ \hline
$^{127}$I & 100 & 5/2 & \begin{matrix}0.264\\0.066\end{matrix} & \begin{matrix}0.309\\0.075\end{matrix}  & \begin{matrix}1.515\\0.655\end{matrix} & \begin{matrix}1.338\\0.779\end{matrix} & 2.74 & 2.775 & 2.8133 & \cite{Ressell:1997kx}\\ \hline
$^{129}$Xe & 26.4 & 1/2 & \begin{matrix}0.007\\0.248\end{matrix} & \begin{matrix}0.01\\0.329\end{matrix}  & \begin{matrix}0.274\\0.03\end{matrix} & \begin{matrix}0.372\\-0.185\end{matrix} & 0.636 & -0.72 & -0.778 & \cite{Menendez:2012tm},\cite{Ressell:1997kx}\\ \hline
$^{131}$Xe & 21.2 & 3/2 & \begin{matrix}0.005\\0.199\end{matrix} & \begin{matrix}-0.009\\-0.272\end{matrix} & \begin{matrix}0.284\\1.419\end{matrix} & \begin{matrix}0.165\\1.572\end{matrix}  & 1.016 & 0.86 & 0.6919 & \cite{Menendez:2012tm}, \cite{Ressell:1997kx} \\
 \end{tabular}
 \caption{Spin and angular momentum matrix elements and magnetic moments for isotopes with non-zero spin, as deduced from the nuclear response functions of \cite{Fitzpatrick:2012ix,Anand:2013yka} at $y=0$  (``th'' for ``theory'') or as given by the most sophisticated calculation in the literature (``lit'').  Natural abundance (NA) and total angular momentum ($J$) are also included. Ref.~\cite{Menendez:2012tm} does not report the orbital angular momentum matrix element (though it does provide the magnetic moment). However Ref.~\cite{Ressell:1997kx} provides $ \LN $ for xenon isotopes as well as iodine, for two different models (so-called ``Bonn A'' (BA) and ``Nijmegen II'' (NII)). We have reported $ \LN $ from \cite{Ressell:1997kx} for the model that is closest to the spin matrix values of \cite{Menendez:2012tm} (BA for $^{131}$Xe and NII for $^{129}$Xe). For iodine, we report the BA model values, because BA comes closest to the experimental value of the magnetic moment. }\label{tab: spins etc}
\end{table}

\begin{table}
\begin{tabular}{r r l r l r l r l r l}
 			& \multicolumn{2}{c}{~~~~Fluorine~~~~}	& \multicolumn{2}{c}{~~~~Sodium~~~~}	& \multicolumn{2}{c}{~~~~Germanium~~~~}	& \multicolumn{2}{c}{~~~~Iodine~~~~}	& \multicolumn{2}{c}{~~~~Xenon~~~~} \\
		$A$=	& \multicolumn{2}{c}{19}				& \multicolumn{2}{c}{23}				& \multicolumn{2}{c}{70,72,73,74,76}		& \multicolumn{2}{c}{127}				& \multicolumn{2}{c}{128-132,134,136} \\
 	$(N,N')=$	& ~~$(p,p)$ 		& $(n,n)$			& ~~$(p,p)$ 		& $(n,n)$			& ~~$(p,p)$ 		& $(n,n)$				& ~~$(p,p)$ 		& $(n,n)$			& $(p,p)$ 		& $(n,n)$ \\
 $\W_M^{(N,N')}(0)$	
 			&~~81			&100			& ~~121			& 144			& ~~1024			& 1658				&~~2809			&5476			& {\color{red}2911}		& {\color{blue} 5984}	\\
 $\W_{\Sigma'}^{(N,N')}(0)$
 			&~~{\color{red}1.81}	&$< 10^{-3}$		& ~~ 0.273		& 0.002			& ~~$< 10^{-3}$	& 0.057				&~~0.26			&0.016			& $< 10^{-3}$	& {\color{blue}0.168}	\\
 $\W_{\Sigma''}^{(N,N')}(0)$
 			&~~{\color{red} 0.903}&$< 10^{-3}$		& ~~0.136			& $< 10^{-3}$		& ~~$< 10^{-3}$	&  0.029				&~~0.13			&0.008			& $< 10^{-3}$	& {\color{blue}0.084}	\\
 $\W_{\Delta}^{(N,N')}(0)$ 	
 			&~~0.025			&0.018			& ~~0.231			& 0.029			& ~~$< 10^{-3}$	& {\color{blue}0.231}		&~~{\color{red}0.536}	&0.100		& 0.015		& 0.119	\\
 $\W_{\Phi''}^{(N,N')}(0)$	
 			&~~0.039			&0.255			& ~~1.48			& 2.43 			& ~~45.3			& 15.4				&~~{\color{red}201}	&44.4			& 117		& {\color{blue}202}	 
 \end{tabular}
 \caption{ Natural-abundance-weighted nuclear response functions at $y=0$ for various target nuclei. Nuclear response functions were evaluated using the code described in \cite{Anand:2013yka}. The target with the largest effective response for neutrons (blue) or protons (red) is highlighted in each row. }\label{tab: numerical responses}
\end{table}

Not only can the relative strengths of the novel nuclear responses be different than that of the standard responses from target to target, but also the behavior of the responses as a function of momentum transfer can be different.  For larger nuclei---the nuclei least well modeled as point particles---the spin- and/or orbital- angular-momentum-dependent form factors can have quite different dependence on energy than each other and than that of the spin-independent/Helm form factors. We plot these dependences explicitly in Appendix~\ref{sec: form factors}, where we refer the reader for details. As is well known, the form factors matter more for larger nuclei, and especially for larger nuclei with non-zero spin, but even more so if momentum/velocity- dependent interactions are involved so that $\Delta$ and $\Phi''$ can be relevant.

In order to get a sense for the importance of the potentially novel momentum dependence of the novel form factors, we consider rates given the interactions described in Sec.~\ref{approach to models} using both the ``full'' form factors as provided by \cite{Fitzpatrick:2012ix,Anand:2013yka}, and using an approximate spin-independent or spin-dependent ``foil'' form factor. For anapole and magnetic dipole DM we replace the combination of nuclear responses in the square brackets of \eqref{eq: anapole rate} and \eqref{eq: mag dip rate} with its value at $y=0$ times the nuclear charge form factor, 
\beq
F^2_\text{E}(y) \equiv {\W^{(p,p)}_M(y) \over \W^{(p,p)}_M(0)}, \label{eq: SI form factor foil}
\eeq
so that the entire rate is proportional to this form factor. 
For $(\vec{L}\cdot \vec{S})$-generating DM, we replace $\W_{\Phi''}^{(N,N')} + {C_\chi \over 4} {m_N^2 \over m_\chi^2} \W_{\Sigma'}^{(N,N')}$ with its value at $y=0$, times the fermi form factor,
\beq
F_\text{SI}^{(N,N')}(y) \equiv {\W^{(N,N')}_M(y) \over \W^{(N,N')}_M(0)}.\label{eq: SI form factor foil LS}
\eeq

For pseudoscalar-mediated spin-dependent DM, as a foil we replace $\W_{\Sigma''}$ with its value at $y=0$, times the spin-dependent response,
\beq
F_\text{SD}^{(N,N')}(y) = {\W_{\Sigma'}^{(N,N')}(y)+\W_{\Sigma''}^{(N,N')}(y) \over \W_{\Sigma'}^{(N,N')}(0)+\W_{\Sigma''}^{(N,N')}(0) }. \label{eq: SD form factor foil}
\eeq
We choose the sum of $\Sigma'$ and $\Sigma''$ as the foil because this combination is usually quoted for the standard spin-dependent form factors utilized in the literature.

\begin{figure}
\includegraphics{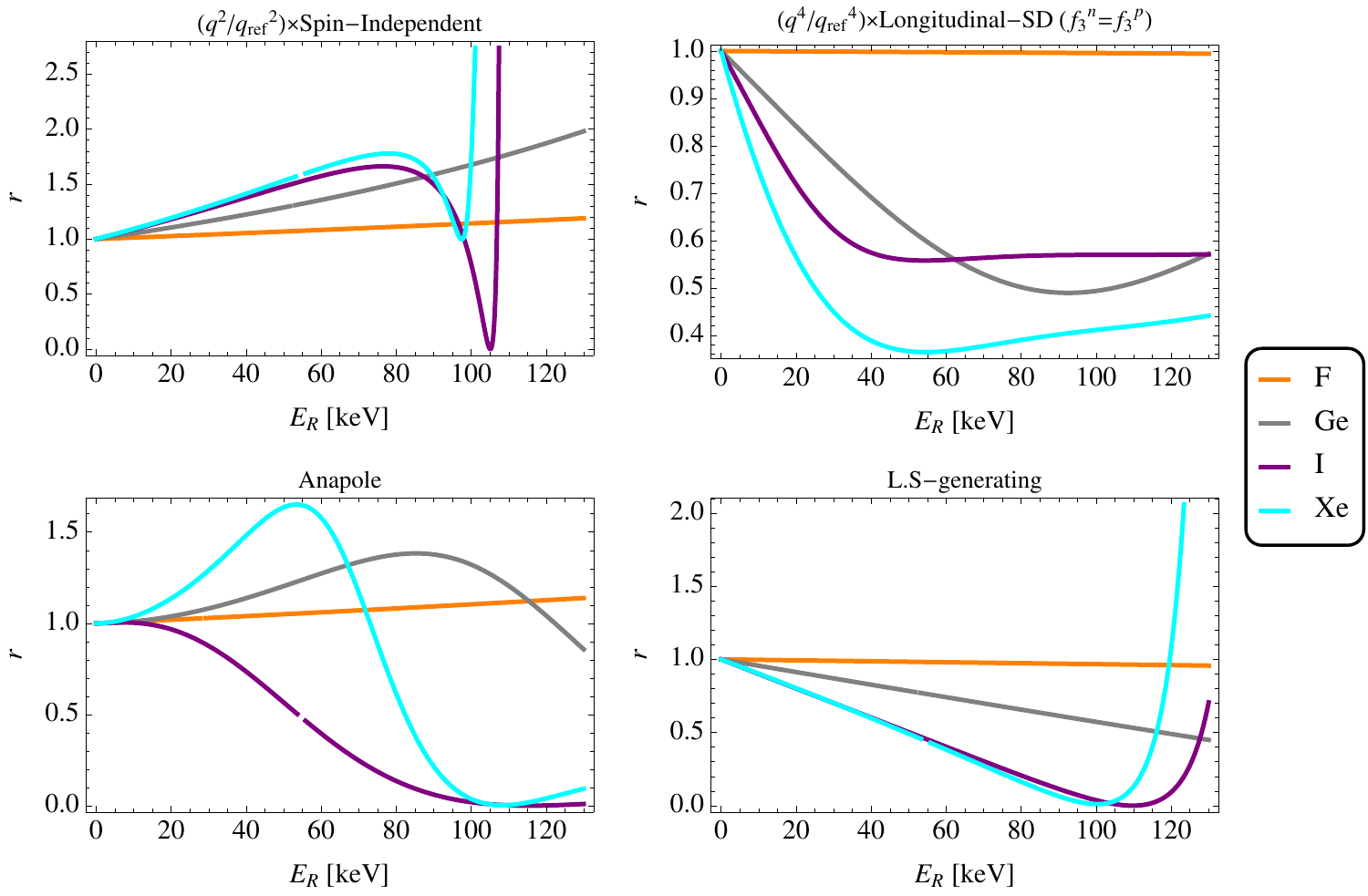}
\caption{Ratio, $r$, of ``foil'' rate to full rate, which is equivalent to the ratio of foil to full form factors for all but the Anapole case. See discussion in the text. 
}\label{fig: ratio plot}
\end{figure}

The ratio, $r$, of the rate as computed using the above ``foil'' form factors to the rate with the full form factors is shown in Fig.~\ref{fig: ratio plot} for our benchmark scenarios. The rate ratio,  $r$, is independent of DM mass for $(\vec{L}\cdot \vec{S})$-generating \eqref{eq: LS rate},  $q^4$-suppressed longitudinal-spin-dependent, and $q^2$-suppressed spin-independent interactions \eqref{eq: pseudoscalar rate} because in these cases the ratio of rates is just a ratio of form factors. The magnetic dipole \eqref{eq: mag dip rate} ratio is extremely similar to that of the anapole  \eqref{eq: anapole rate}, and $r$ is very mildly mass dependent in these cases because the rate is a mass-dependent linear combination of novel and spin-independent form factors.  To illustrate the difference between foil and full rates for the anapole and magnetic dipole, in Fig.~\ref{fig: ratio plot} we show the ratio of anapole rates for $m_\chi=100$ GeV. The foil form factor used in the $q^2\times$spin-independent ratio is the Helm form factor, and should be used as a guide for estimating ``typical'' error associated with the momentum-dependence of form factors.

Significant (order 50\%) differences between the Helm and full SI form factors of \cite{Fitzpatrick:2012ix,Anand:2013yka} arise for the heaviest elements at recoil energies of order 50 keV, which, even for a $q^2$-suppressed SI interaction, sits on the tail of the differential rate as a function of recoil energy for order 100 GeV DM  (see Fig.~\ref{fig: q2si fit}). As DM mass and/or momentum suppression associated with the underlying interaction increases, we expect the error due to momentum dependence of form factors to be more significant due to the fact that in such cases the tail of the differential rate moves to higher energies. In other words, given larger DM masses or more significant momentum suppression in the DM-nucleon scattering rate, a larger portion of the total scattering rate comes from higher momentum transfer events, for which the amount of rate suppression due to the form factors is less certain. The uncertainty associated with form factor rate suppression could be especially important for, {\em e.g.}, threshold-based bubble chamber experiments like COUPP with a heavy target like iodine because this is precisely the kind of experiment where a high proportion of higher-momentum-transfer events could be contributing to the total rate. On the other hand, the recoil energy range probed by current xenon-target experiments cuts off before 50 keV, so we should expect uncertainty in form factor suppression to be less important in interpreting the results of such experiments. (These statements are made concrete with constraints from existing and simulated data in Sec.~\ref{sec: bounds}.) 

We should also keep in mind that the tail of the differential rate is affected by (and sometimes \emph{controlled} by, especially for light DM and light target elements) the tail of the DM velocity distribution, which has its own associated uncertainties. Generally speaking, we should expect uncertainties in the velocity distribution to be more important for light DM than for heavy DM because a higher proportion of direct detection scattering events are likely to originate from DM at the tail of the velocity distribution, where uncertainties are greatest.  The effect of moderate changes to the velocity distribution ({\em e.g.} lowering the escape velocity, adding streams, or using a non-Maxwellian distribution) on light DM constraints and ROIs was explored in \cite{Gresham:2013mua} and \cite{Fitzpatrick:2010br},\footnote{See also \cite{Green:2010gw,Green:2011bv}.} and it was found that such changes had little effect on constraints and ROIs for a variety of targets and underlying interactions, including, {\em e.g.}, anapole and magnetic dipole interactions.\footnote{An exception can be annual modulation experiments, for which a stream can make a more dramatic difference \cite{Savage:2006qr}.} As a reference point for comparison with the foil to full rate ratios, the fractional change in the velocity moment $g(v_\text{min}) = g (\sqrt{2 m_T E_R}/2 \mu_T)$, which is the astrophysics dependent part of the differential rate \cite{Fox:2010bz,Gondolo:2012rs,DelNobile:2013cva}, given a SHM distribution with $v_0=220$ km/s and $v_{esc} = 544$ km/s versus either a SHM distribution with $v_\text{esc} = 490$ km/s or the non-Maxwellian distribution of \cite{Bhattacharjee:2012xm},  is only order 10\% when the moment has dropped to 10\% of its value at $E_R=0$.  

Note that, for xenon and iodine, the SI form factor falls off to zero near 100 keV. Near this recoil energy the ratio of the Helm to the Ref.~\cite{Fitzpatrick:2012ix,Anand:2013yka} SI form factor asymptotes to infinity because the Ref.~\cite{Fitzpatrick:2012ix,Anand:2013yka} form factor hits zero first (see Appendix \ref{sec: form factors}); of course since the total rate will be very small near 100 keV, the difference between the Helm and novel form factors will be of little consequence in this energy range. On the other hand, some of the novel form factors for xenon and iodine fall off much slower than the SI form factors so that the rate out closer to 100 keV can be more important;\footnote{This phenomena has also been noted for the standard SD form factor \cite{Bednyakov:2006ux}.}  we will see that this is the case for, {\em e.g.}, momentum-suppressed spin-dependent and anapole DM. As demonstrated by our benchmark models, cases in which novel responses arise tend to be precisely the cases in which the differential rate can be weighted towards larger momentum transfer. Thus it could be important to understand the behavior of novel form factors out to larger recoil energies.

In the next section we more concretely explore the effect of form factors on the interpretation of DM direct detection experiments by looking at the effects of the response functions on the constraints extracted from existing experiments.  We also simulate data from a hypothetical experiment with a heavy DM candidate and see how the response functions affect the ROIs inferred from the data.

\subsection{The effect of form factor momentum dependence on the interpretation of direct detection experiments}\label{sec: fits and bounds}

\subsubsection{The effect of form factors on fits to simulated data}\label{sec: fits}

To make the relative importance of including proper form factors clear, we simulate DM scattering events on fluorine, germanium, iodine, and xenon targets, and then fit the data given different underlying assumptions about the form factors relevant for the interaction.  One hundred events were generated assuming an underlying distribution for 80 GeV or 250 GeV DM scattering via a representative set of the benchmark interactions discussed in \S\ref{approach to models} with the full form factors. Perfect resolution and acceptance are assumed.  For fits, exposures were adjusted so that 100 events would be expected in the 0-100 keV energy range at the same cross section that leads to 100 events off of iodine given an exposure of $10^5$ kg-days and $m_\chi=250$ GeV. For comparison, we also simulated 100 events in the narrower 0-50 keV recoil energy range, and exposures were similarly adjusted for fits. We fit the data using either the proper form factors or the ``foil'' form factors discussed in \S\ref{sec: momentum dependence of form factors}. Binned log likelihood ($\ln L$) was computed for 10 keV bins given the 0-100 keV range or 5 keV bins given the 0-50 keV range. Region-of-interest contours are set using $\ln L = \ln L_\text{max}- \text{CDF}^{-1}(\text{ChiSq}[2], \text{C.L.}) / 2$ with C.L.=68\% .

Here we aim to concretely demonstrate how a reasonable yet in-principle-inaccurate model of the momentum dependence of the nuclear response of a target can affect an inference of the underlying WIMP physics. We have modeled our analysis on idealized experiments that can measure the energy of scattering events with very good resolution.\footnote{For example, we have not modeled our fluorine target ``experiment'' after a more realistic bubble chamber experiment,  which is sensitive only to energy thresholds rather than to the absolute energy of individual scattering events. See \cite{Peter:2013aha} for an approach to the inverse problem that takes account of the different direct detection technologies.}

\begin{figure}
\includegraphics{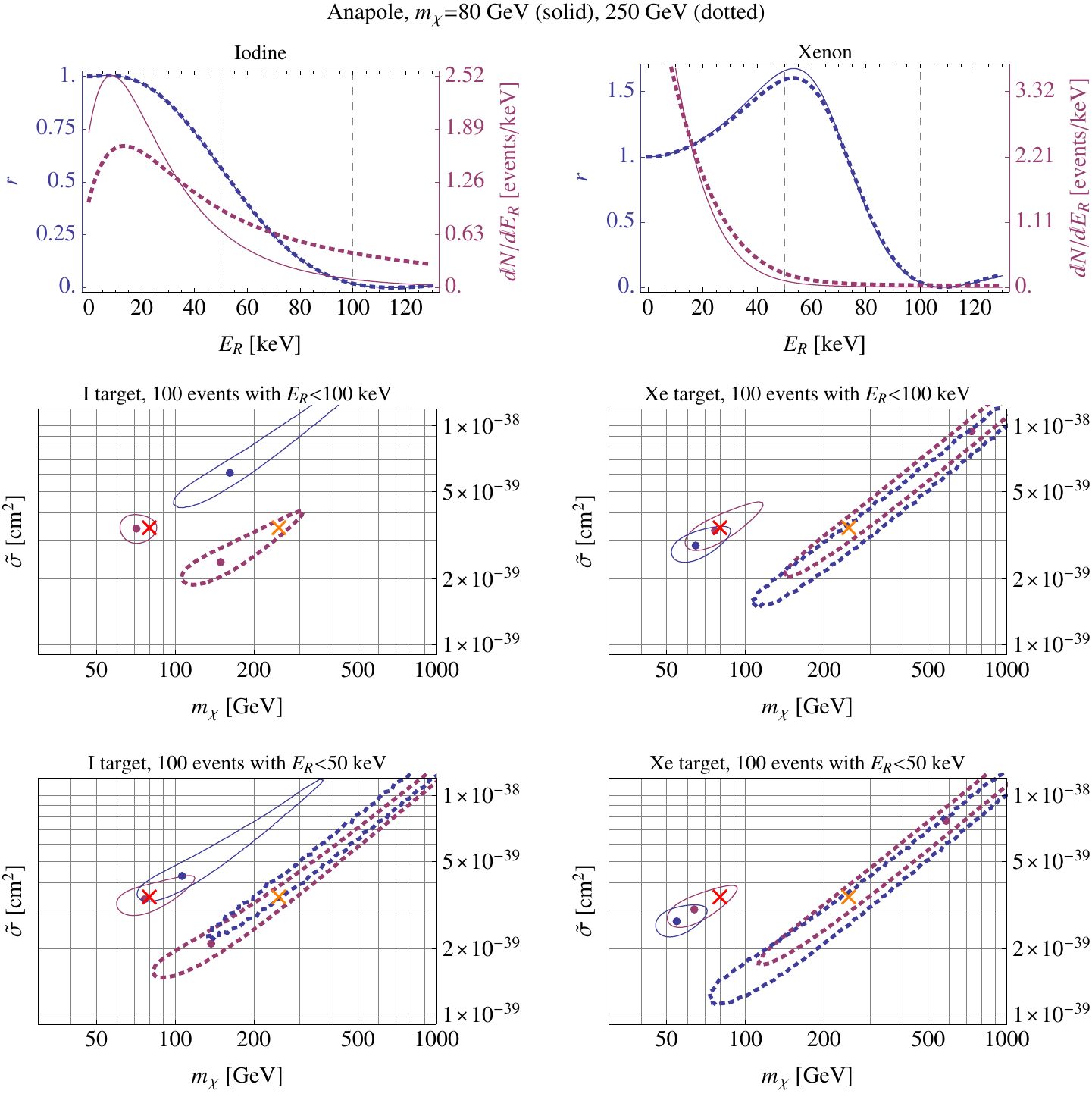}
\caption{For the anapole interaction, (top two panels) expected event spectrum (pink) alongside the ratio of the foil rate to the true rate (blue), and (bottom four panels) fits for idealized iodine-target and xenon-target experiments assuming full form factors (pink, used to generate the events in the first place) or foil form factors (blue). True mass and cross sections are marked with an ``$\times$.''  The solid is for simulated 80 GeV DM and the dashed for 250 GeV.  In the middle left panel no curve appears for the 250 GeV case because a fit with the wrong form factors gives a poor fit to the data. The results from fits to two sets of simulated data (100 events with $0<E_R<50$ keV or $0<E_R<100$ keV) for each target are shown in the bottom four panels.}\label{fig: small AP panel}
\end{figure}

\begin{figure}
\includegraphics{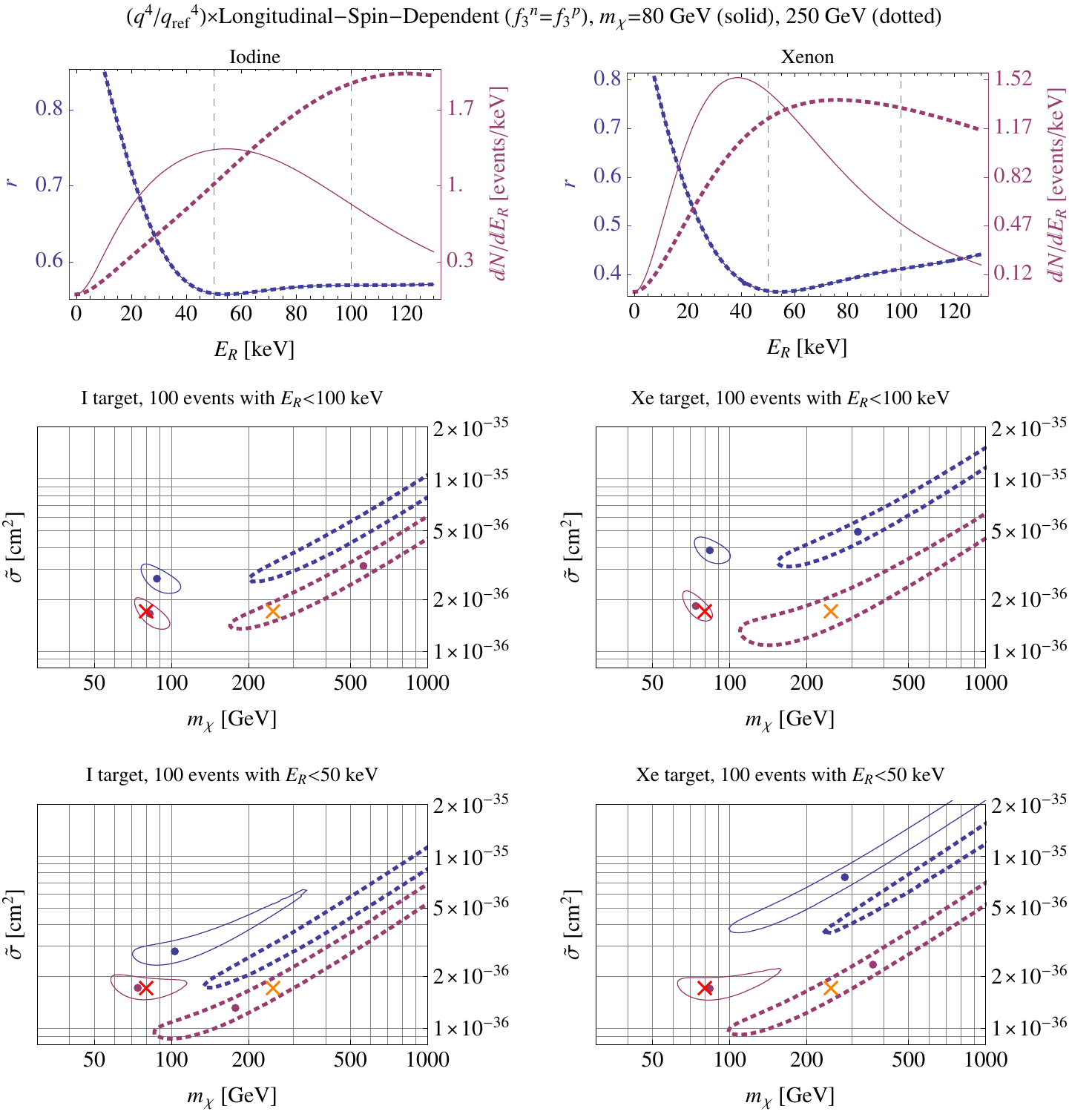}
\caption{For a momentum-dependent longitudinal spin-dependent interaction, (top two panels) expected event spectrum (pink) alongside the ratio of the foil rate to the true rate (blue), and (bottom four panels) fits for idealized iodine-target and xenon-target experiments assuming full form factors (pink, used to generate the events in the first place) or foil form factors (blue). True mass and cross sections are marked with an ``$\times$.''  The solid is for simulated 80 GeV DM and the dashed for 250 GeV.  The results from fits to two sets of simulated data (100 events with $0<E_R<50$ keV or $0<E_R<100$ keV) for each target are shown in the bottom four panels.}  \label{fig: small q4sd panel}
\end{figure}

In Figs.~\ref{fig: small AP panel} and \ref{fig: small q4sd panel} we show some examples for which novel form factors have the most dramatic effect upon the interpretation of simulated events for the benchmark models we examined. A comprehensive set of plots for all benchmark models can be found in Appendix \ref{sec: simulated data plots}. In each figure, the top panels show the spectrum of expected events given full form factors (which were used to simulate events), alongside the ratio of the rate given foil form factors to full form factors (c.f.~Fig.~\ref{fig: ratio plot}).  The lower four panels show the results of fits to simulated data, taking the energy range of 0-50 keV and 0-100 keV, in order to see the effect of the higher-energy recoil events on the fits.  

Given an anapole interaction, the scattering rate off of iodine over the range of energies with significant rate  has a shape that is substantially affected by the nuclear form factor (and even more so for 250 GeV DM than for 80 GeV DM).  Thus the interpretation of a preferred mass range given anapole scattering is fairly dependent on choosing the correct form factor (see Fig.~\ref{fig: small AP panel}). For simulations with a  0-50 keV energy range, the foil form factor fits are better than in the 0-100 keV range case because the shape of the full and foil form factors differs most in the 50$+$ keV recoil energy range, and the overall rate is also substantial in this range. 

In Fig.~\ref{fig: small q4sd panel}, we see that the form factor suppression for large elements like iodine and xenon is quite different for the standard spin-dependent case versus a pseudoscalar-mediated scenario in which only the longitudinal component of spin contributes. Thus the inferred WIMP-nucleon cross section (and less-so the WIMP mass range) can be quite different if a standard spin-dependent form factor is assumed versus a longitudinal-spin-dependent form factor.   In this case the shape of the foil and full form factors differs substantially below 50 keV, and from 50 to 100 keV the ratio levels out; at the same time the differential rates peak around 50 keV for 80 GeV DM and above 100 keV for 250 GeV DM, so the 50-100 keV range is weighted heavily.  Consequently the foil form factor fits for the 0-50 keV energy range simulations predict a higher-than-actual cross section and a skewed-from-actual mass range (especially for 250 GeV DM). The fits for the 0-100 keV range give a fairly accurate and precise mass range and (again) a  higher-than-actual cross section. The difference in form factors basically manifests as an overall normalization difference in the 0-100 keV range case. 

The take-away lesson here is that momentum-suppressed interactions such as the anapole and pseudoscalar-mediated interactions lead to larger momentum transfer events being weighted more heavily, making them more sensitive to the momentum dependence of nuclear responses; this is particularly relevant for large elements such as iodine and xenon. Similarly, rates given larger DM masses are weighted higher at larger momentum transfer, so again, form factors are more relevant for higher-mass DM.

\subsubsection{Update of bounds from current experiments for benchmark models}\label{sec: bounds}

Lastly, we show the effect of novel form factors upon the interpretation of some representative contemporary null direct detection experiments. Updated bounds for our benchmark models from LUX and XENON100, CDMS II, COUPP, and PICASSO are shown in Fig.~\ref{fig: bounds}. See Appendix \ref{sec: experiments} for details. We include these representative xenon, germanium, iodine-fluorine(-carbon), and fluorine target experiments to show (a) the complementarity of the different targets in setting bounds given a larger swath of possible interactions in which novel nuclear responses arise and (b) for which targets bounds are most affected by form factors. (We include XENON100 in addition to LUX because the energy range probed by XENON100 ($\sim$7-45 keV) covers substantially larger energy than that of LUX ($\sim$4-25 keV).)  In most cases, the bounds are not highly affected by the momentum dependence of form factors, except to some extent for COUPP. This is because COUPP is a bubble chamber threshold energy experiment with an iodine target. COUPP is sensitive to both (i) events on the tails of form factor distributions, by virtue of its large target, and (ii) a higher proportion of potentially large-momentum-transfer events, by virtue of the fact that the experiment is sensitive to the rate integrated from some threshold energy up to infinite energy. (See also the discussion in Sec.~\ref{sec: momentum dependence of form factors}.) Xenon-target experiments share feature (i) with COUPP but not feature (ii); in contrast the highest recoil energies probed by current xenon-target experiments are only order 40 keV and thus are minimally affected by uncertainties in or novel behavior of form factors at large momentum transfer.

\begin{figure}
\includegraphics{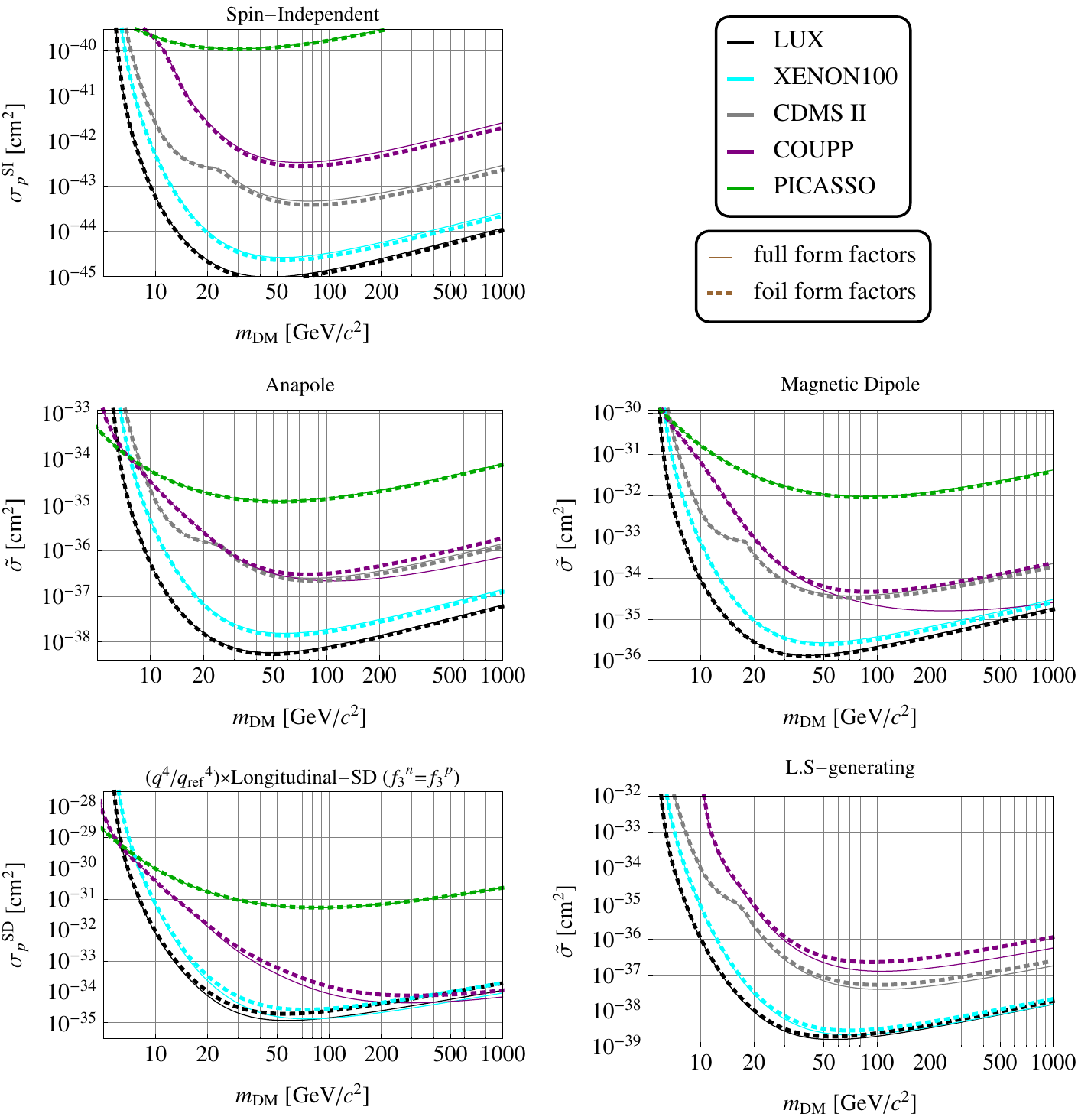}
\caption{Constraints on DM-nucleon cross sections from LUX and XENON100 (Xe targets), CDMS II (Ge Target),  COUPP (F and I target), and PICASSO (F target) for scattering via our benchmark models. Solid lines show constraints using the form factors provided by \cite{Fitzpatrick:2012ix,Anand:2013yka} and dashed were derived assuming the foil form factors discussed in Sec.~\ref{sec: momentum dependence of form factors}. The spin-independent constraints given by \cite{Fitzpatrick:2012ix,Anand:2013yka} or Helm form factors are shown for reference.}\label{fig: bounds}
\end{figure}

\section{Conclusions}\label{sec: conclusions}

In the context of a set of UV-complete benchmark models for which novel nuclear responses dominate over the standard spin-independent or -dependent responses (see Sec.~\ref{approach to models}), we have provided concrete demonstrations of the importance of novel nuclear form factors that can arise in well-motivated nonstandard scenarios \cite{Fitzpatrick:2012ix}. Some of the effects of nonstandard nuclear responses have already been captured in past treatments that did not employ the nuclear response language of \cite{Fitzpatrick:2012ix}. For example for light DM, both the anapole and dipole interactions are adequately captured by previous treatments.  
On the other hand, as the momentum transfer is increased, the effect of  the new momentum dependence of the new nuclear responses becomes important, and the standard form factors cannot be safely used as substitutes. This new momentum dependence is most important for heavy elements such as iodine and xenon with abundant isotopes that have an unpaired nucleon, in momentum-/velocity-dependent interactions such as anapole and dipole interactions.  They are also important for momentum-suppressed spin-dependent interactions, for which a different combination of two independent spin responses enters as compared to the standard spin-dependent case. In general, it is more important to understand the momentum dependence of form factors---including novel form factors---when the underlying nucleon-DM interactions are momentum suppressed and therefore the rates are weighted towards larger recoil energy. We also demonstrated that, the larger the energy range probed by an experiment, the more relevant the form factors become. The momentum dependence of novel nuclear responses for smaller elements such as germanium, and for yet smaller elements like fluorine and sodium, is practically negligible over the recoil energy range relevant for direct detection.  However, $^{73}$Ge is very sensitive to the orbital-angular-momentum response, so if a germanium-based experiment were to probe an order 100+ keV energy range, it could be important to have better theoretical control over the orbital angular momentum form factor. 

As the nature of the weak scale DM becomes increasingly constrained, the types of nonstandard types of interactions we have focused on here will continue to be the source of theoretical study.  Detection of DM will require a broad set of tools and theories in order to uncover its nature, and the application of nuclear physics to DM detection is crucial for correctly modeling this behavior.  Here we have offered concrete examples, tools and practical advice for the DM theorist as we continue to broaden the scope of models constrained or discovered.

\acknowledgments
We thank Nikhil Anand and Liam Fitzpatrick for correspondence and clarification of some of their work.  The work of KZ is supported by NASA astrophysics theory grant NNX11AI17G and by NSF CAREER award PHY 1049896.

\appendix

\section{Nuclear response ``coefficients''}\label{sec: response coefficients}

We provide, for completeness of our discussion, expressions for the nuclear response coefficients $R_k$, as provided originally in \cite{Anand:2013yka}. The ``coefficients'' are functions of the nucleon-WIMP operator coefficients $c_i^N$ as well as WIMP velocity and momentum transfer.

\begin{align}
 R_{M}^{N N'}(\vec{v}_T^{\perp 2}, {\vec{q}^{\,2} \over m_N^2}) &= c_1^N c_1^{N^\prime } + {C_\chi \over 4} \left[ {\vec{q}^{\,2} \over m_N^2} \vec{v}_T^{\perp 2} c_5^N c_5^{N^\prime }+\vec{v}_T^{\perp 2}c_8^N c_8^{N^\prime }
+ {\vec{q}^{\,2} \over m_N^2} c_{11}^N c_{11}^{N^\prime } \right] \nonumber \\
 R_{\Phi^{\prime \prime}}^{N N'}(\vec{v}_T^{\perp 2}, {\vec{q}^{\,2} \over m_N^2}) &= {\vec{q}^{\,2} \over 4 m_N^2} c_3^N c_3^{N^\prime } + {C_\chi \over 16} \left( c_{12}^N-{\vec{q}^{\,2} \over m_N^2} c_{15}^N\right) \left( c_{12}^{N^\prime }-{\vec{q}^{\,2} \over m_N^2}c_{15}^{N^\prime} \right)  \nonumber \\
 R_{\Phi^{\prime \prime} M}^{N N'}(\vec{v}_T^{\perp 2}, {\vec{q}^{\,2} \over m_N^2}) &=  c_3^N c_1^{N^\prime } + {C_\chi \over 4} \left( c_{12}^N -{\vec{q}^{\,2} \over m_N^2} c_{15}^N \right) c_{11}^{N^\prime } \nonumber \\
  R_{\tilde{\Phi}^\prime}^{N N'}(\vec{v}_T^{\perp 2}, {\vec{q}^{\,2} \over m_N^2}) &={C_\chi \over 16} \left[ c_{12}^N c_{12}^{N^\prime }+{\vec{q}^{\,2} \over m_N^2}  c_{13}^N c_{13}^{N^\prime}  \right] \nonumber \\
   R_{\Sigma^{\prime \prime}}^{N N'}(\vec{v}_T^{\perp 2}, {\vec{q}^{\,2} \over m_N^2})  &={\vec{q}^{\,2} \over 4 m_N^2} c_{10}^N  c_{10}^{N^\prime } +
  {C_\chi \over 16} \left[ \left(c_4^N+{\vec{q}^{\,2} \over m_N^2}c_6^N \right)\left(c_4^{N^\prime}+{\vec{q}^{\,2} \over m_N^2}c_6^{N^\prime} \right) + \right.  \nonumber \\
 & \left. +\vec{v}_T^{\perp 2} c_{12}^N c_{12}^{N^\prime }+{\vec{q}^{\,2} \over m_N^2} \vec{v}_T^{\perp 2} c_{13}^N c_{13}^{N^\prime } \right] \nonumber \\
    R_{\Sigma^\prime}^{N N'}(\vec{v}_T^{\perp 2}, {\vec{q}^{\,2} \over m_N^2})  &={1 \over 8} \left[ {\vec{q}^{\,2} \over  m_N^2}  \vec{v}_T^{\perp 2} c_{3}^N  c_{3}^{N^\prime } + \vec{v}_T^{\perp 2}  c_{7}^N  c_{7}^{N^\prime }  \right]
       + {C_\chi \over 16} \left[ c_4^N c_4^{N^\prime} +  \right.\nonumber \\
       &\left. {\vec{q}^{\,2} \over m_N^2} c_9^N c_9^{N^\prime }+{\vec{v}_T^{\perp 2} \over 2} \left(c_{12}^N-{\vec{q}^{\,2} \over m_N^2}c_{15}^N \right) \left( c_{12}^{N^\prime }-{\vec{q}^{\,2} \over m_N^2}c_{15}^{N^\prime} \right) +{\vec{q}^{\,2} \over 2 m_N^2} \vec{v}_T^{\perp 2}  c_{14}^N c_{14}^{N^\prime } \right] \nonumber \\
     R_{\Delta}^{N N'}(\vec{v}_T^{\perp 2}, {\vec{q}^{\,2} \over m_N^2})&=  {C_\chi \over 4} \left[ {\vec{q}^{\,2} \over m_N^2} c_{5}^N c_{5}^{N^\prime }+ c_{8}^N c_{8}^{N^\prime } \right] \nonumber \\
 R_{\Delta \Sigma^\prime}^{N N^\prime}(\vec{v}_T^{\perp 2}, {\vec{q}^{\,2} \over m_N^2})&= {C_\chi \over 4} \left[c_{5}^N c_{4}^{N^\prime }-c_8^N c_9^{N^\prime} \right].
 \label{eq: response coefficients}
\end{align}

\section{Form factors}\label{sec: form factors}

\begin{figure}
\includegraphics{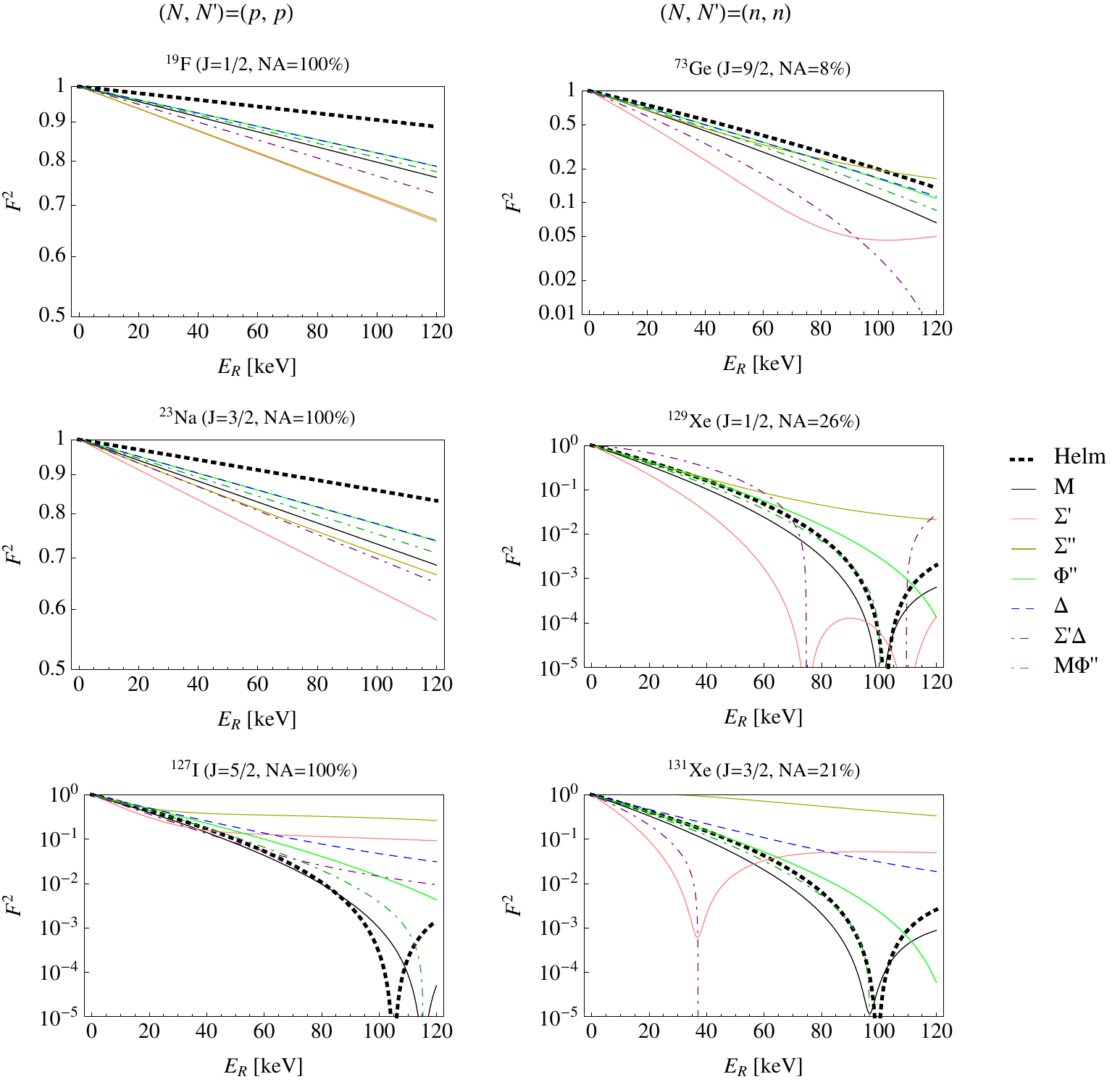}
\caption{ Form factors \eqref{eq: form factor definition} with $X=M,\Sigma',\Sigma'',\ldots$ as indicated in the legend and $(N,N')=(p,p)$ (left) or $(n,n)$ (right) for several target nuclei with nonzero spin, alongside the Helm form factor.}\label{tab: numerical form factors}
\end{figure}

\begin{figure}
\includegraphics{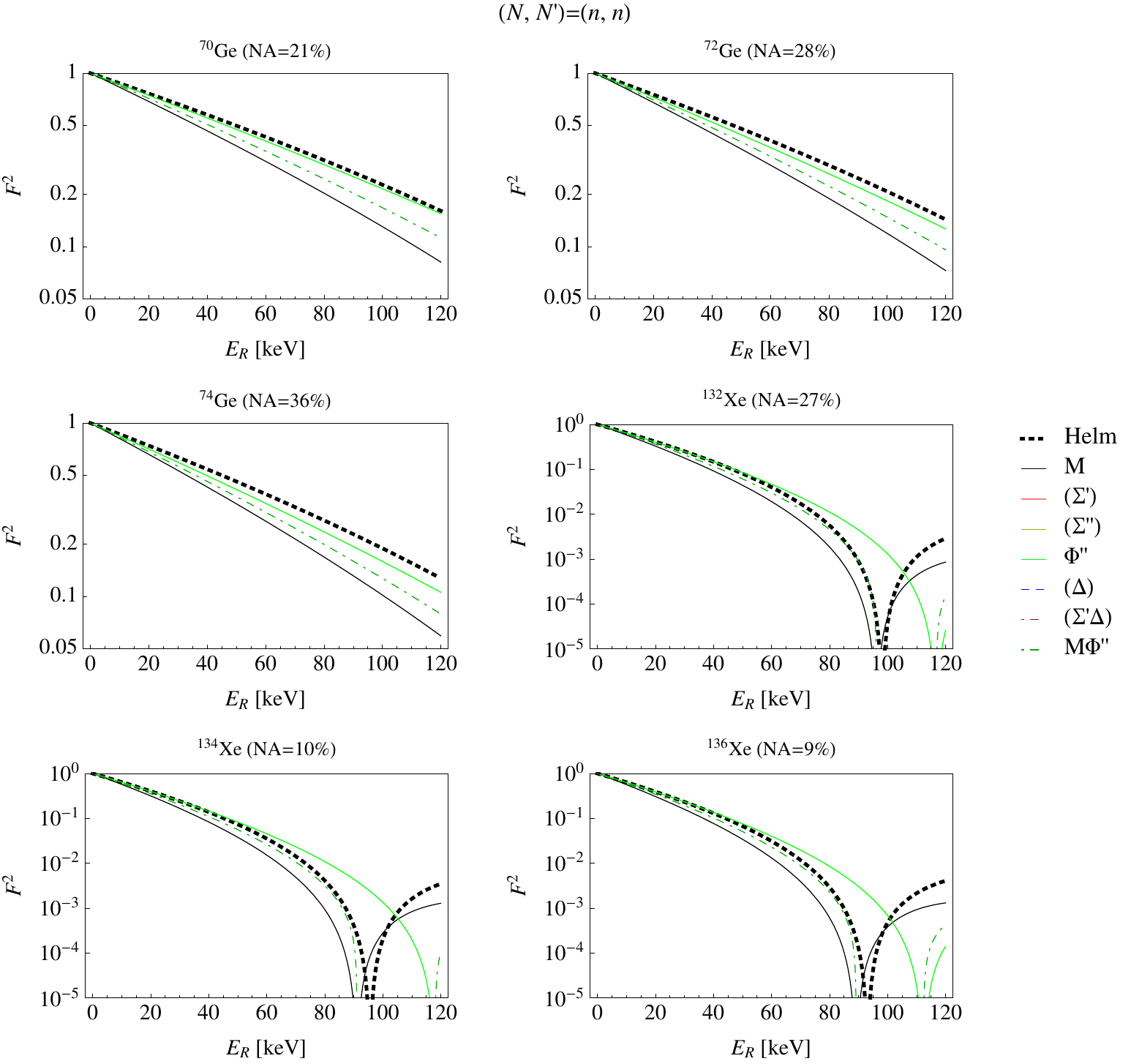}
\caption{Form factors \eqref{eq: form factor definition} with $X=M,\Sigma',\Sigma'',\ldots$ as indicated in the legend and $(N,N')=(n,n)$ for zero-spin germanium and xenon isotopes, alongside the Helm form factor.}\label{tab: numerical form factors no spin}
\end{figure}

As a convenient reference for determining the momentum dependence of novel form factors (especially in comparison to standard form factors), in Figs.~\ref{tab: numerical form factors} and \ref{tab: numerical form factors no spin} we show form factors normalized to one at $q^2=0$, as calculated from the code of \cite{Anand:2013yka}. More specifically, we show normalized form factors \beq F^2(E_R) \equiv {W_X^{(N,N')}(y) \over W_X^{(N,N')}(0)} \qquad y=q^2 b^2 / 4 = 2 m_T E_R b^2 / 4, \label{eq: form factor definition} \eeq for $(N,N')=(p,p)$ or $(n,n)$ according to whether the relevant isotopes have mostly unpaired protons or neutrons, for various $X$. The Helm form factor is also shown. The meaning and names of the associated responses are provided in Table \ref{tab: responses} and their relative magnitudes can be read off of Table \ref{tab: numerical responses}. 

\section{Scattering rates and fits to simulated data}\label{sec: simulated data plots}

\begin{figure}
\includegraphics{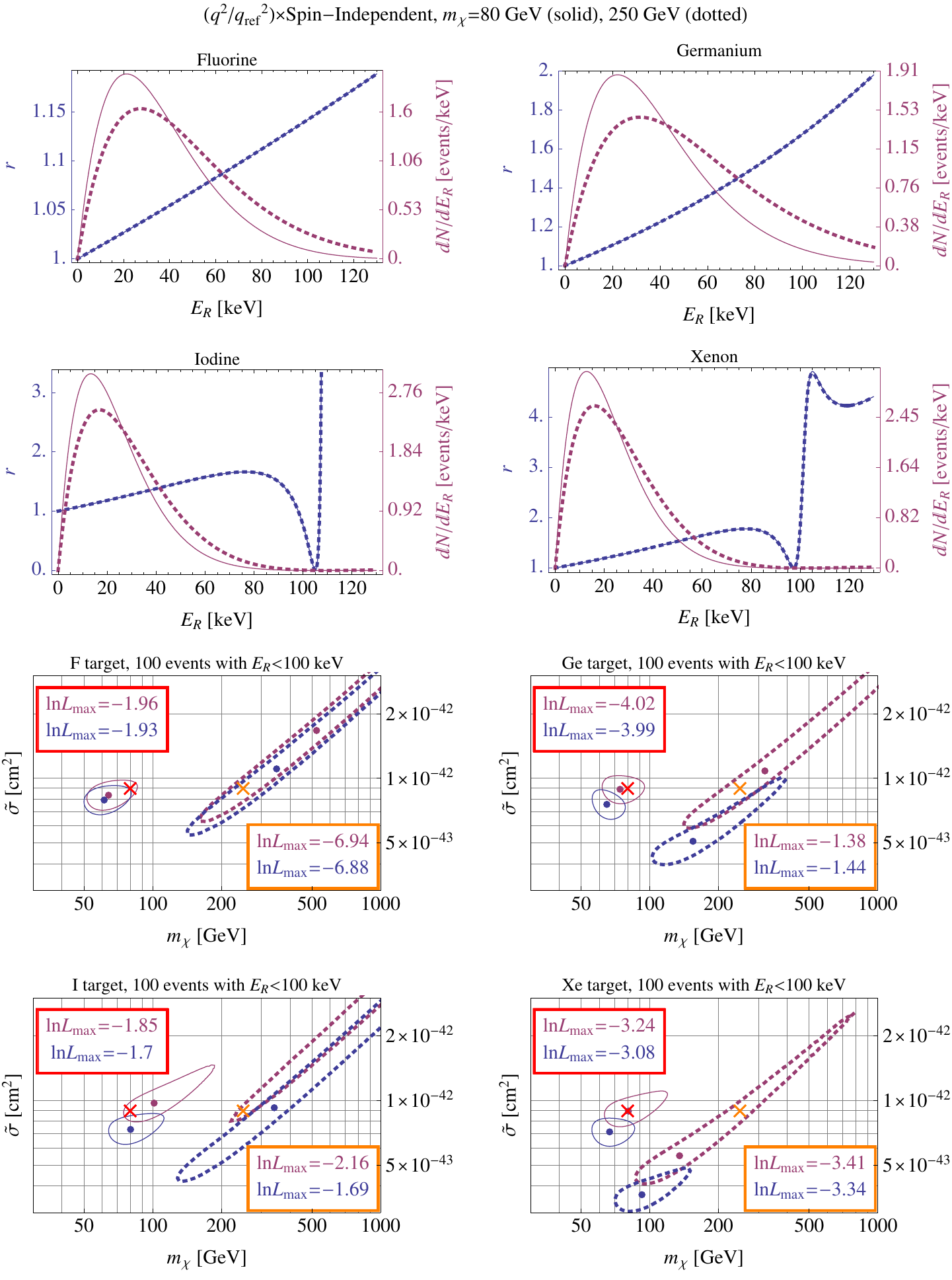}
\caption{Differential rates (\emph{top})  and  68\% C.L. fit contours (\emph{bottom}). Refer to Appendix~\ref{sec: simulated data plots} and the discussion surrounding Figs.~\ref{fig: small AP panel},~\ref{fig: small q4sd panel}.}\label{fig: q2si fit}
\end{figure}

\begin{figure}
\includegraphics{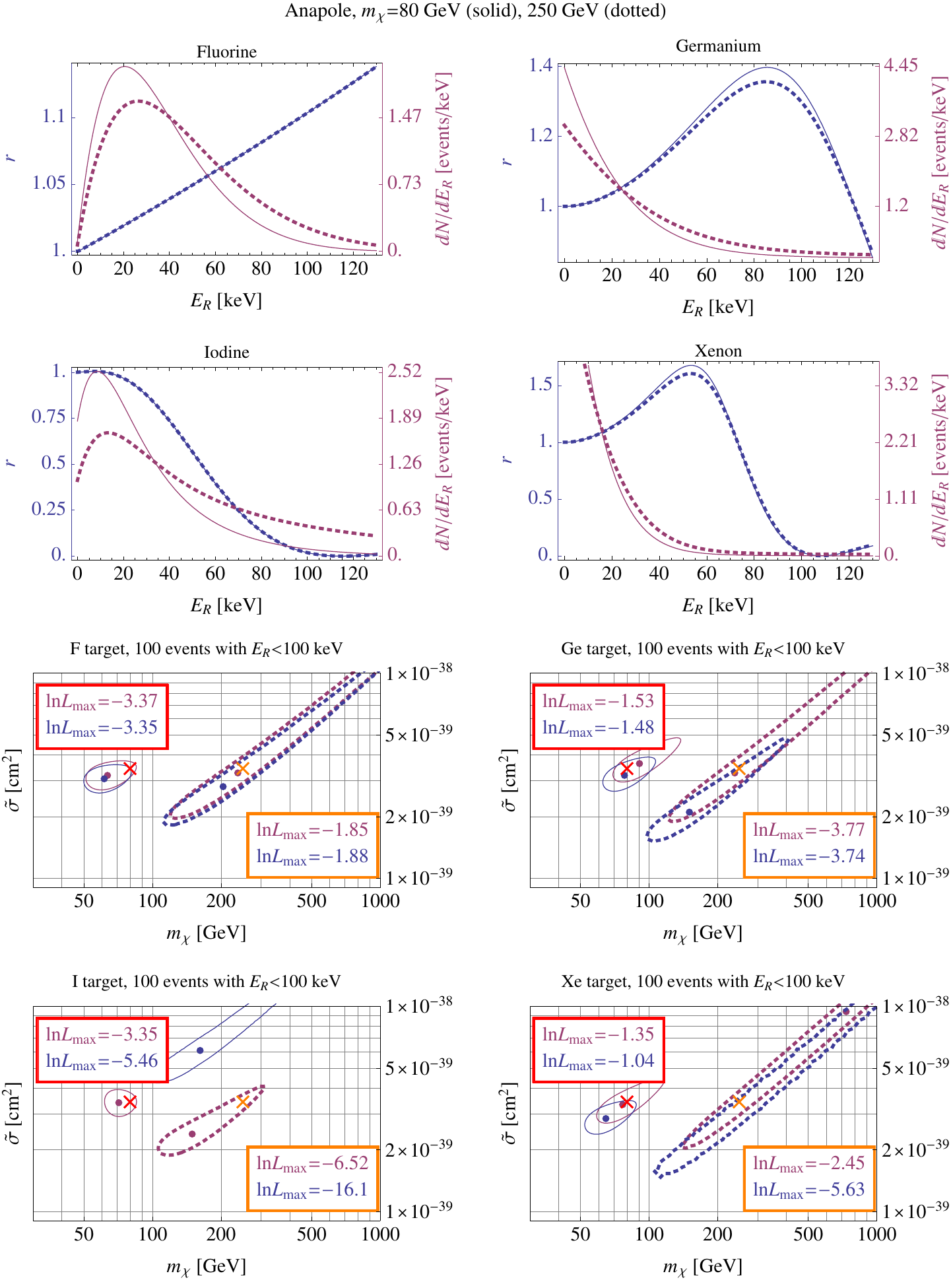}
\caption{Differential rates (\emph{top})  and  68\% C.L. fit contours (\emph{bottom}). For iodine given $m_\chi=250$ GeV, $\ln L_\text{max}$ occurs at order $10^4$ GeV. Refer to Appendix~\ref{sec: simulated data plots} and the discussion surrounding Figs.~\ref{fig: small AP panel},~\ref{fig: small q4sd panel}.}\label{fig: AP fit}
\end{figure}

\begin{figure}
\includegraphics{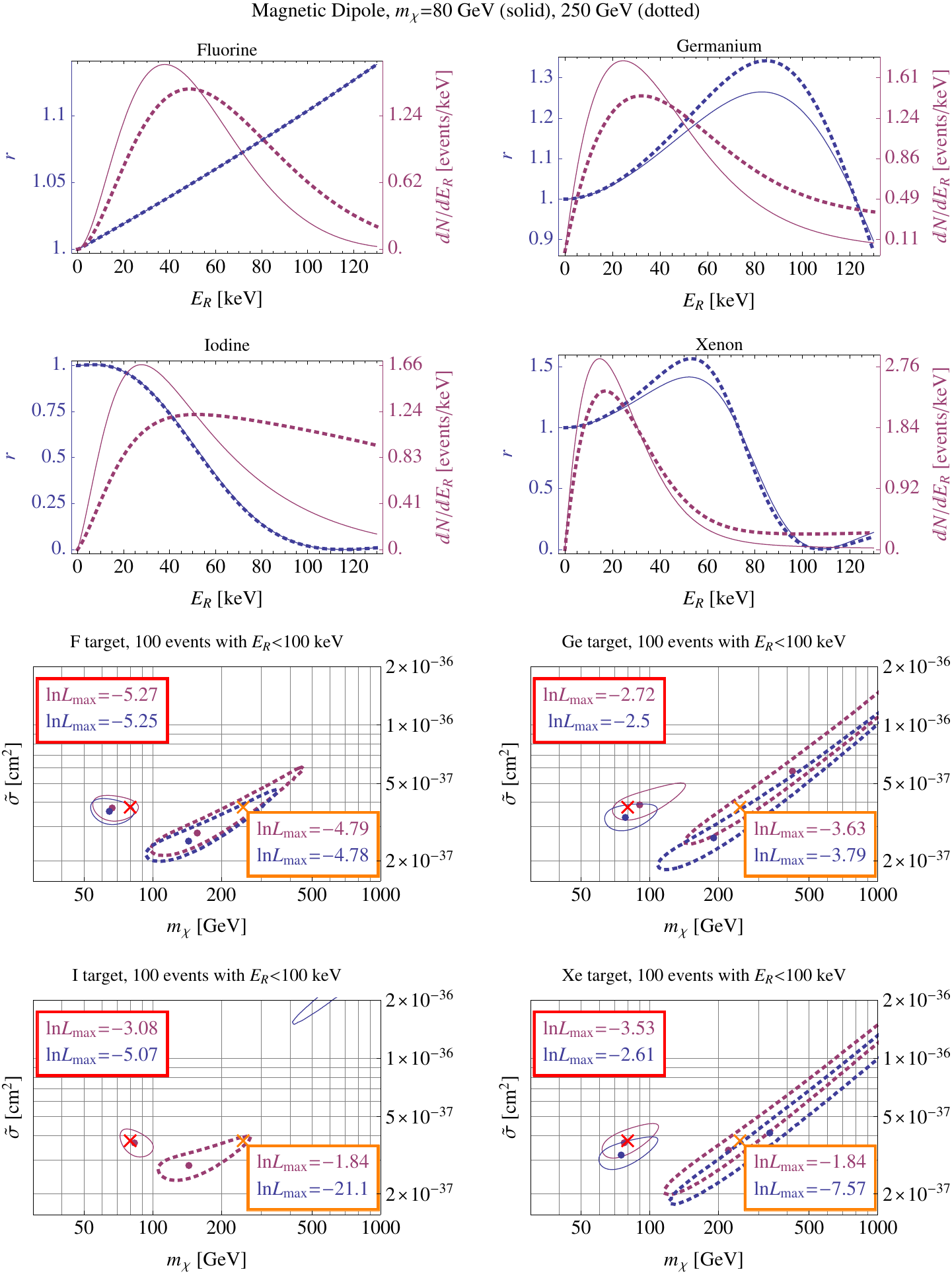}
\caption{Differential rates (\emph{top})  and  68\% C.L. fit contours (\emph{bottom}). For iodine given $m_\chi=250$ GeV, $\ln L_\text{max}$ occurs at order $10^4$ GeV. Refer to Appendix~\ref{sec: simulated data plots}, Fig.~\ref{fig: AP fit} and the discussion surrounding Figs.~\ref{fig: small AP panel},~\ref{fig: small q4sd panel}.}\label{fig: MM fit}
\end{figure}

\begin{figure}
\includegraphics{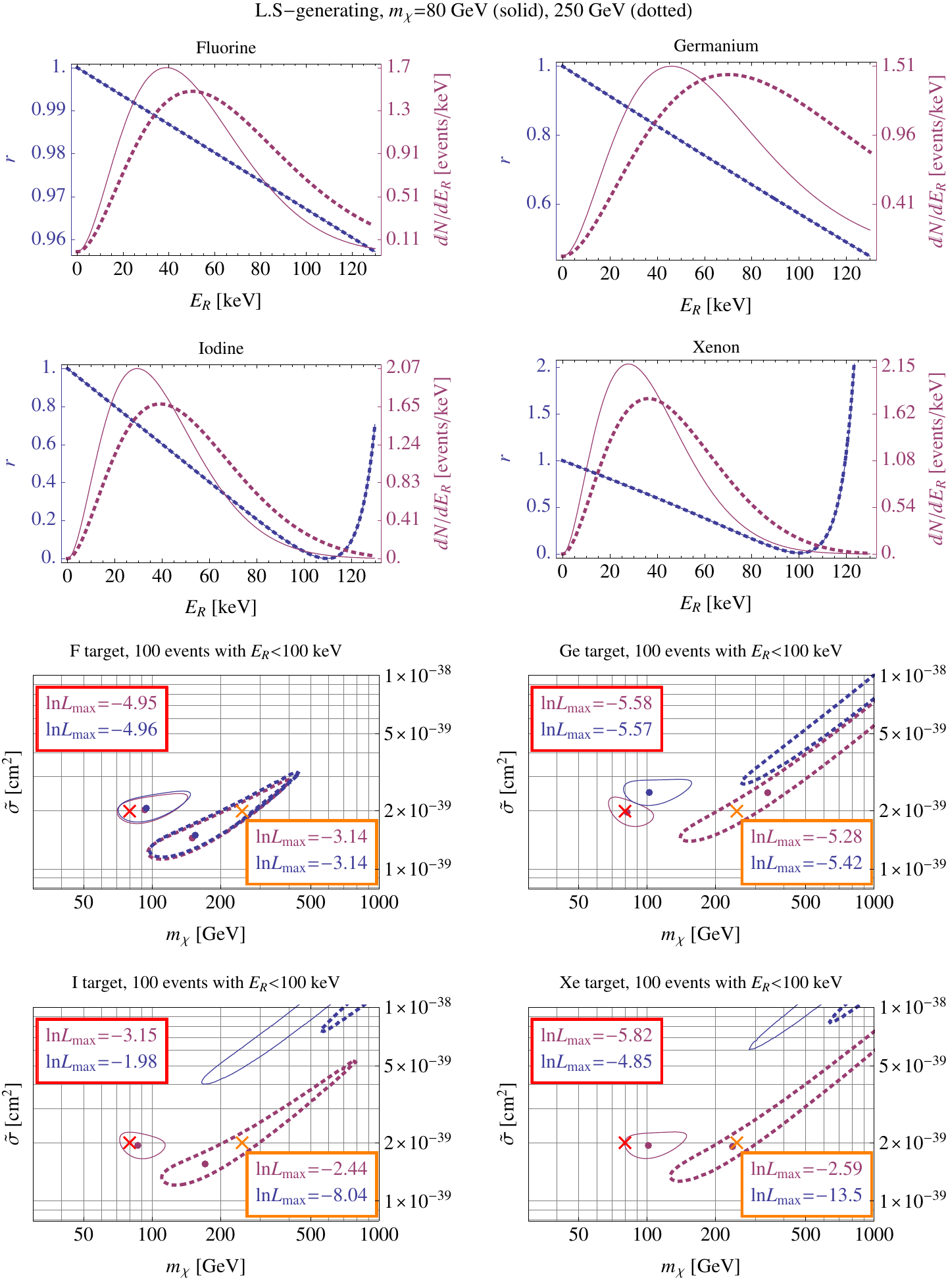}
\caption{Differential rates (\emph{top})  and  68\% C.L. fit contours (\emph{bottom}).  Refer to Appendix~\ref{sec: simulated data plots}, Fig.~\ref{fig: AP fit} and the discussion surrounding Figs.~\ref{fig: small AP panel},~\ref{fig: small q4sd panel}.}\label{fig: LdotS fit}
\end{figure}

\begin{figure}
\includegraphics{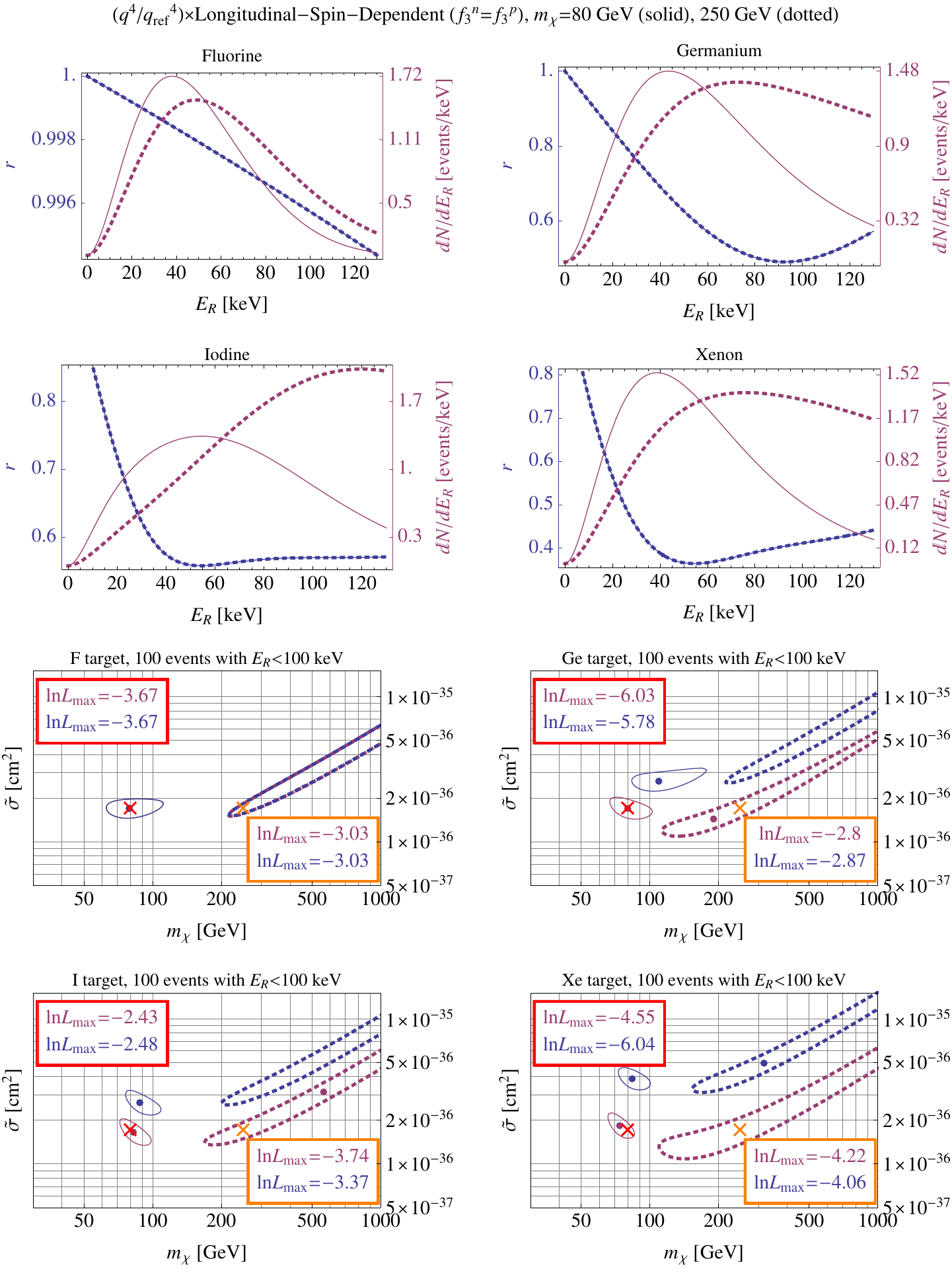}
\caption{Differential rates (\emph{top})  and  68\% C.L. fit contours (\emph{bottom}). Refer to Appendix~\ref{sec: simulated data plots}, Fig.~\ref{fig: AP fit} and the discussion surrounding Figs.~\ref{fig: small AP panel},~\ref{fig: small q4sd panel}.}\label{fig: q4sd fit}
\end{figure}

As a complement to the selected results shown in Figs.~\ref{fig: small AP panel} and \ref{fig: small q4sd panel} we show rates and simulated data for the complete sets of operators (five in total) considered in Sec.~\ref{approach to models} in Figs.~\ref{fig: q2si fit}--\ref{fig: q4sd fit}.  In each figure, the event spectra (\emph{top})  and  68\% C.L. fit contours (\emph{bottom}) with predicted rates employing either the full form factors (red) or foil form factors (blue) are shown. (See Sec,~\ref{sec: momentum dependence of form factors} for a description of foil form factors.) In the spectrum plots, blue curves indicate the ratio of foil to full (c.f. Fig.~\ref{fig: ratio plot}). The red/orange ``$\times$'' marks the true value. The $m_\chi$ range from 10 to $10^4$ GeV was scanned; $\ln L_\text{max}$ in this range are indicated. See Sec.~\ref{sec: fits} for a fuller discussion of the methods used in simulating the data.

\section{Experiments and details for event rates}\label{sec: experiments}

For constraint or region of interest plots presented in Secs.~\ref{sec: light DM} and \ref{sec: bounds}, we follow the same procedure for the experiments shown as detailed in the appendix of \cite{Gresham:2013mua}, though in \ref{sec: bounds} we include an analysis of all CDMS-II Germanium data, which was not included in \cite{Gresham:2013mua}. Additionally, we extend the XENON100 maximum gap analysis to include the photo-electron signal range from 3 to 30 rather than from 3 to 20. In Table \ref{summary experiments}, the nuclear target(s), exposure, and analysis signal range are summarized for a set of current experiments that together have the best potential to constrain a large variety of elastic scattering models.

\def\used{\em}

\begin{table}
\begin{tabular}{l l l l l}
 			& $T$	& Ex 	  		& Ref. 				& keVnr Range   \\
\hline	
CDMS Si$^{*,\dagger}$ 		& Si 		& 140.2 kg-days 	& \cite{Agnese:2013rvf} 	& 7-100 \\
CRESST-II$^{*,\dagger}$	& O,Ca,W	& 730 kg-days	& \cite{Angloher:2011uu} & $\sim10/Q_X$-$300\footnote{The analysis range ends at 40 keV, but data well separated from background are shown up to 300 keV.}/Q_X$)\footnote{Quenching factors $Q_X$ for $X=$O,Ca,W are estimated to be about 0.1, 0.06, and 0.04, respectively.}\\
DAMA$^\dagger$ 		& Na,I 	& 1.17 ton-yr 		& \cite{Bernabei:2010mq} & 6.7-67\footnote{For sodium, assuming a quenching factor $Q_\text{Na}=0.3$.}  \\
CoGeNT$^\dagger$ 		& Ge 	& 266 kg-days 		& \cite{Aalseth:2012if} 	& 2.3-11	\\
CDMS II		& Ge		& 974 kg-days		& \cite{Ahmed:2008eu,Ahmed:2009zw}	& 10-100 \\
CDMSlite$^*$		& Ge		& 6 kg-days		& \cite{Agnese:2013lua}	& 0.84-24 \\
CDMS Ge L-E 	& Ge 	& 35 kg-days\footnote{Includes only detector T1Z5, which is the most constraining.} 
									& \cite{Ahmed:2010wy} 	& 2-100  \\
Xenon10 S2$^*$ 	& Xe 	& 15 kg-days 		& \cite{Angle:2011th} 	& 1.4-10\\
XENON100 	& Xe 	& 7636 kg-days 	& \cite{Aprile:2012nq} 	& 6.6-44 \\
LUX			& Xe		& 10065 kg-days 	& \cite{Akerib:2013tjd}	& 3.6-24.8	 \\
PICASSO		& F		& 114.3 kg-days	& \cite{Archambault:2012pm} 	& thresholds from: 1.7-55	\\
COUPP		& C,F,I	& 437.4 kg-days\footnote{After cuts.}	 
									 & \cite{Behnke:2012ys} 		& thresholds from: 7.8-15.5\\
\end{tabular}
\caption{Experiments/analyses considered in this work along with a few other experiments that could be competitive in setting limits giving a large swath of possibilities for elastic scattering. A $^*$ indicates that we do not explicitly show results for this experiment in this paper, and a $^\dagger$ indicates that the experiment reports a possible signal. We also include the target ($T$), total exposure (before cuts), the primary reference, and recoil energy range (in keVnr). The nuclear recoil energy range quoted is the average expected energy corresponding to the signal range boundaries, so, generally speaking, energies on tails of distributions beyond this range are probed.}\label{summary experiments}
\end{table}

CDMS II results from two different sets of runs (123-124 and 125-128) are detailed in \cite{Ahmed:2008eu} and \cite{Ahmed:2009zw}.  Two events with energies 12.3 and 15.5 keV were observed in the second set of runs and none in the first. We use the maximum gap method to set 90\% C.L. limits based on the two events observed in runs 123-125. We approximate resolution as being perfect. We digitize the efficiency as a function of energy for each set of runs (See Fig.~6.23 of \cite{Ahmed:2011osa}) and take the effective runs 123-128 efficiency to be an exposure-weighted sum of the two efficiencies. The exposure of runs 123-124 is reported as 397.8 kg-days in \cite{Ahmed:2008eu}, but we take it to be 9\% lower than this based on the statement in \cite{Ahmed:2009zw}. We take the exposure for runs 125-128 to be 612.13 kg-days. 


\bibliography{../direct_detection}

\end{document}